\documentclass{aa}  

\usepackage{subcaption}
\usepackage{graphicx}
\usepackage{txfonts}
\usepackage{amsmath}
\usepackage{caption}
\usepackage{subcaption}

\usepackage{lscape}
\usepackage{tocloft}
\usepackage{natbib}
\usepackage{hyperref}
\graphicspath{{./}{}}

\begin{document}

\title{Period-Luminosity Relations, projection factor and radii of Anomalous Cepheids}
\titlerunning{PL relations, projection factor and radii  of Anomalous Cepheids}
\authorrunning{Wielg\'orski, Pietrzyński, Gieren et al.}

\author{P. Wielg\'orski
\inst{1}\thanks{\email{pwielgor@camk.edu.pl}}
\and
G. Pietrzy\'nski
\inst{1,2}
\and
W. Gieren
\inst{2}
\and
B. Zgirski
\inst{2}
\and
W. Narloch
\inst{1}
\and
G. Hajdu
\inst{1}
\and
J. Storm
\inst{3}
\and
N. Nardetto
\inst{4}
\and
P. Kervella
\inst{5}
\and
B. Pilecki
\inst{1}
\and
M. G\'orski
\inst{1}
\and
R. Smolec
\inst{1}
\and
R. Salinas
\inst{1}
\and
D. Graczyk
\inst{1}
\and
V. Hocd\'e
\inst{4}
\and
P. Karczmarek
\inst{1}
\and
M. Taormina
\inst{1}
\and
W. Pych
\inst{1}
\and
H. Netzel
\inst{1}
\and
R. Chini
\inst{1,6}
\and
K. Hodapp
\inst{7}
\and
M. Ka\l{}uszy\'nski
\inst{1}
\and
F. Pozo Nu\~nez
\inst{8}
\and
K. Kotysz
\inst{9}
\and
D. Mo\'zdzierski
\inst{9}
\and
P. Miko\l{}ajczyk
\inst{9}
\and
P. Ko\l{}aczek-Szyma\'nski
\inst{9,10}
}

\institute{Nicolaus Copernicus Astronomical Center, Polish Academy of Sciences, Bartycka 18, 00-716 Warszawa, Poland
\and
Universidad de Concepci\'on, Departamento de Astronom\'{i}a, Casilla 160-C, Concepci\'on, Chile
\and
Leibniz-Institut f\"ur Astrophysik Potsdam (AIP), An der Sternwarte 16, 14482 Potsdam, Germany
\and
Université Côte d'Azur, Observatoire de la Côte d'Azur, CNRS, Laboratoire Lagrange, France
\and
LESIA, Observatoire de Paris, Universit\'e PSL, CNRS, Sorbonne Universit\'e, Universit\'e Paris Cit\'e, 5 place Jules Janssen, 92195 Meudon,France
\and
Astronomisches Institut, Ruhr-Universit\"at Bochum, Universit\"atsstrasse 150, D-44801 Bochum, Germany
\and
University of Hawaii, Institute for Astronomy, 640 N. Aohoku Place, Hilo, HI 96720, USA
\and
Astroinformatics, Heidelberg Institute for Theoretical Studies, Schloss-Wolfsbrunnenweg 35, 69118 Heidelberg, Germany
\and
Astronomical Institute, University of Wroc\l{}aw, M. Kopernika 11, 51–622 Wroc\l{}aw, Poland
\and
Warsaw University Observatory, Al. Ujazdowskie 4, 00-478 Warszawa, Poland
}

\date{Received ..., 2024; accepted ...., 2024}

\abstract
    {Anomalous Cepheids are radially pulsating stars observed in dwarf galaxies, the Galactic bulge and halo, and globular clusters. Similarly to other radially pulsating stars, they can be used as distance indicators through their Period$-$Luminosity Relations (PLRs) and the geometrical Baade$-$Wesselink (BW) method.}
    {We aim to calibrate the zero$-$point of the distance scale of Anomalous Cepheids using nearby representatives of this class of pulsating stars.}
    {We collected optical and near$-$infrared photometry and spectra for a sample of nearby Anomalous Cepheids with two telescopes located at the $Rolf$ $Chini$ Cerro Murphy Observatory (former Observatory Cerro Armazones) and optical telescopes offered by the Las Cumbres Observatory, and with instruments hosted and operated by the European Southern Observatory. Using parallaxes measured by the $Gaia$ space mission and mean magnitudes from our new photometry, we calibrate the zero$-$point of the PLRs in Johnson $B$, $V$, 2MASS $J$, $H$, $K_S$, and Pan$-$STARRS $g$, $r$, $i$ passbands and selected Wesenheit indices. Slopes are adopted from the PLRs of the Large Magellanic Cloud Anomalous Cepheids from the literature. Using the surface brightness$-$colour relation version of the BW technique, we also determined the projection factors and mean radii of three nearby Anomalous Cepheids.}
    {Precision of the measured zero$-$points is at the level of 0.04$-$0.05mag and their systematic uncertainty is estimated to about 0.1mag. We used our zero$-$points and literature photometry of the Large Magellanic Cloud Anomalous Cepheids to measure the distance modulus of this galaxy and obtained a value of 18.454$\pm$0.045(statistical)\,mag, in a very good agreement with the most accurate value from eclipsing binaries. The obtained projection factors are 1.38$\pm$0.13, 1.59$\pm$0.21 and 1.35$\pm$0.14 for V716~Oph, XX~Vir and UY~Eri, respectively. We suspect that the very high projection factor of XX~Vir is the result of the underestimated parallax of this star in the $Gaia$ catalogue as its radius and absolute brightness are also higher than expected for Anomalous Cepheids at this period. The radii measured for V716~Oph and UY~Eri are in agreement with the period-radius relation obtained from the Large Magellanic Cloud Anomalous Cepheids.}
    {}

\keywords{solar neighbourhood $-$-- Stars: distances $-$-- Stars: variables: Cepheids $-$-- Magellanic Clouds}
\maketitle

\section{Introduction} \label{sec:intro}

Anomalous Cepheids (ACeps, also known as BL Boo type stars) are low metallicity radially pulsating stars ($[Fe/H]<-$1.5dex) observed mostly in dwarf galaxies, the Galactic bulge, the halo and rarely in globular clusters \citep[see e.g.][and references therein]{2008AcA....58..293S,2022AJ....164..191N,2024AJ....167..247B,2024AA...682A...1R}. Periods of known ACeps span a range of 0.6 and 2.5 days for fundamental mode pulsators and 0.4, and 1.2 days for first overtone pulsators. On the Period$-$Luminonosity diagram they are located between short period Classical and Type II Cepheids. Pulsational masses are about 1.2$-$2$M_{\odot}$ \citep[e.g.][]{1997AJ....113.2209B,2004AA...417.1101M,2012AA...540A.102F,2017AA...603A..70G,2024AA...682A...1R}. \citet{2017ApJ...842..110P} measured the dynamical mass of a pulsating star with a period of about 4 days which lies on the extension of the Period$-$Luminosity Relation (PLR, the Leavitt Law) of ACeps in the Large Magellanic Cloud (LMC), being a member of a binary system. They obtained ~1.5$M_{\odot}$, however, the classification of this star is not clear. There are two hypotheses explaining the origin of Anomalous Cepheids: they can be intermediate ($\sim$5Gyr) age stars with exceptionally low metal abundance or the result of the mass transfer in an old population binary system, crossing the Instabillity Strip of Cepheids during the core helium burning phase \citep[e.g.][and references therein]{2012AA...540A.102F,2024AA...682A...1R,2018AcA....68..213I}. 

The vast majority of known ACeps was discovered in the course of the Optical Gravitational Lensing Experiment \citep[OGLE, ][]{2015AcA....65....1U} in the Large and Small Magellanic Clouds and Galactic bulge \citep{2008AcA....58..293S,2017AcA....67..297S,2018AcA....68..315U}. Optical and near$-$infrared observations of these stars performed by several projects allowed the construction of the PLRs for ACeps that can be used as standard candles for measuring distances of clusters and nearby galaxies \citep[e.g.][]{2025AA...699A.370S}. There are very few ACeps known in the general field \citep{2019AA...625A..14R,2024AA...682A...1R,2018SerAJ.197...13J} and before the $Gaia$ space mission \citep{2016A&A...595A...1G} the classification of those stars was unclear. Accurate parallaxes provided recently by $Gaia$ open a possibility to use nearby ACeps to calibrate the zero$-$point of the distance scale of this class of stars with good accuracy.

Another method which can be used to measure distances of radially pulsating stars like ACeps is the Baade$-$Wesselink technique \citep[BW, also known as the parallax$-$of$-$pulsation, ][]{1926AN....228..359B,1940ZA.....19..249B,1947BAN....10..252W}. Photometry or interferometry is used to measure the angular size variations over the pulsational cycle of the star while from the integration of the radial velocity curve (obtained from time$-$resolved spectra) we get the physical radius displacement of the star. These two quantities from simple geometry give the distance of the star. The bottle$-$neck of this method is the so$-$called projection factor (p$-$factor) which allows to estimate the pulsational velocity of the photosphere from the measured radial velocities. The value of the p$-$factor is primarily defined by the geometrical projection of the pulsational velocity of every part of the stellar disc onto the line of sight, but also by the limb darkening and velocity gradients in the atmosphere \citep[see e.g.][and references therein]{1986PASP...98..881H,2007AA...471..661N}. This makes the theoretical predictions very complicated. The distance measured with the BW techniqe linearly depends on the assumed value of the p$-$factor thus its accurate calibration is a key for the use of the BW techniqe to measure distances with good accuracy. Very precise parallaxes of nearby pulsating stars like Classical, Type II and Anomalous Cepheids and RR Lyrae type stars from $Gaia$ can be used in the inversed BW analysis to calibrate the p$-$factor for different classes of pulsating stars. This will open a possibility to use new forthcoming facilities like, e.g. the Extremely Large Telescope, to measure geometrical distances of Local Group galaxies and constitute the new anchors for the distance scale of Cepheids, the Tip of the Red Giant Branch (TRGB) or the J$-$region Asymptotic Giant Branch (JAGB) used in the distance ladder to determine the Hubble Constant \citep{2024arXiv240806153F,2024ApJ...977..120R}. 

In this work we present the first determination of the PLRs in optical Johnson  $B$, $V$, near$-$infrared Two Micron Sky Survey \citep[2MASS, ][]{2006AJ....131.1163S} $J$, $H$ and $K_{\mathrm{S}}$ and the Panoramic Survey Telescope and Rapid Response System \citep[Pan$-$STARRS][]{2012ApJ...750...99T} $g$, $r$, $i$ bands and selected Wesenheit indices \citep{1982ApJ...253..575M} for Milky Way field Anomalous Cepheids, and the very first BW analysis of the representatives of this class of pulsating stars with the aim to determine their p$-$factor and radii. This work is another in the series of publications of the Araucaria Project \citep{2023arXiv230517247A} in which we present the calibration of methods of measuring distances using various types of pulsating stars using new optical and near$-$infrared photometry from the $Rolf$ $Chini$ Cerro Murphy Observatory (OCM)\footnote{\url{https://ocm.camk.edu.pl/}} and $Gaia$ parallaxes. Our dataset is described in section \ref{sec:data}. Section \ref{sec:analysis} contains our analysis and discussion of the results. The summary of our work is presented in section \ref{sec:summary}.

\section{Data} \label{sec:data}
We selected a sample of candidates for nearby ACeps, Type II Cepheid and RR Lyrae type stars, based on SIMBAD\footnote{\url{https://simbad.cds.unistra.fr/simbad/}} and AAVSO\footnote{\url{https://vsx.aavso.org/}} databases. We limited the sample to the stars visible from northern Chile ($\delta < $+30$^{\circ}$) and we preferred stars with expected distances closer than 5kpc to assure high precision of parallaxes from $Gaia$. The stars confirmed to be Type II Cepheids and RR Lyrae type have been published in \citet{2022ApJ...927...89W} and \citet{2023ApJ...951..114Z}. The list of ACeps candidates for which we managed to collect photometric and spectroscopic data contains 8 stars: XZ~Cet, V338~Pup, FY~Aqr, V716~Oph, BF~Ser, CE~Her, XX~Vir and UY~Eri.

\subsection{Optical and near$-$infrared photometry, reddening and periods}\label{sec:data_mw}

We collected photometric time$-$series data for our sample stars in Johnson $B$, $V$, 2MASS $J$, $H$ and $K_{\mathrm{S}}$, and Pan$-$STARRS $g$, $r$, $i$ passbands. 

Near$-$infrared photometry in the $J$, $H$ and $K_{\mathrm{S}}$ passbands has been collected between March 2017 and March 2020  with the 0.8\,m Infra-Red Imaging System \citep[IRIS,][]{2010SPIE.7735E..1AH} telescope located at OCM. Data reductions, photometry and the standardization process is described in detail in \citet{2022ApJ...927...89W}. Following \citet{2022ApJ...927...89W}, we adopt 0.025mag as the systematic error of our transformations.

Optical photometry in the Johnson $B$ and $V$ passbands have been obtained between March 2017 and March 2020 with 0.4\,m $VYSOS16$ telescope located at OCM \citep{2013AN....334.1115R}. Details about $VYSOS16$ data processing can be found in \citet{2024AA...689A.241W}. The aperture photometry made with a dedicated pipeline based on \texttt{Astropy} Python library \citep{2018AJ....156..123A} and \texttt{DAOPHOT} \citep{1987PASP...99..191S} was transformed to the standard Johnson system using secondary standards from the synthetic allsky catalogue described in \citet{2023AA...674A..33G}, created based on $Gaia$ low resolution spectra. Similarily as in \citet{2024AA...689A.241W}, we adopt 0.02\,mag as the systematic uncertainty of our transformations.

Photometry in $g$, $r$ and $i$ passbands have been obtained between August 2021 and July 2022 with 16 robotic 0.4 metre telescopes of the Las Cumbres Observatory (LCO) Global Telescope Network\footnote{\url{https://lco.global/}} within programs CLN2021B$-$008 and CLN2022A$-$008. The details of instrumental calibrations and photometry can be found in \citet{2023ApJ...953...14N}. We emphasize that our $g$, $r$ and $i$ photometry was standardized with ATLAS All$-$Sky Stellar Reference Catalog version 2 \citep[ATLAS$-$REFCAT2, ][]{2018ApJ...867..105T} stars as secondary standards, which is in the Pan$-$STARRS photometric system \citep{2012ApJ...750...99T}.

We present a sample of our photometric measurements in Table \ref{tab:photometry}. The full version of this Table is available as suplementary material.

Periods of pulsation for V716~Oph and XX~Vir are adopted from \citet{2022MNRAS.516.2095Y}. For the remaining stars we preliminarily adopted periods from the AAVSO Variable Stars Index database. We then varied periods when fitting the light curves to obtain the smallest possible spread of the measurements around the fit. Adopted periods are presented in Table \ref{tab:acep_data_cd}. For V716~Oph and XX~Vir we used observed$-$minus$-$calculated ($O-C$) diagrams for light curves maxima from \citet{2022MNRAS.516.2095Y} to account for period changes and properly phase light curves and radial velocity curve in the BW analysis. As the period changes of those stars are non$-$linear, we fit a 3rd order polynomial to $O-C$. Our phased light curves are presented in Figures \ref{fig:v16_b}$-$\ref{fig:i} in Appendix \ref{ap:fig}.

Mean magnitudes needed for PLRs are calculated from phased light curves by transforming magnitudes to fluxes. These are then fit with $Akima$ spline polynomials \citep{10.1145/321607}, implemented in the \texttt{Python Scipy} package \citep{2020SciPy-NMeth}. This is integrated to obtain its mean value on the intensity scale, which is then converted back onto the magnitude scale. The uncertainty is evaluated in the following Monte Carlo process. From the normal distribution defined by the original measurements and respective photometric errors we draw artificial measurements. For every light curve we calculate the mean value as described above. We repeat this procedure 2000 times and from the histogram of the mean magnitudes  estimate uncertainties of the mean magnitude from the 16th and 84th percentiles. The mean magnitudes in each passband are presented in Table \ref{tab:acep_data}.


Photometry was corrected for the interstellar extinction using values of the color excess based on the \citet{2011ApJ...737..103S} reddening maps. The adopted values of $E(B-V)$ are  presented in Table \ref{tab:acep_data_cd}. Reddening is significant for V716~Oph only, for the rest of the stars it is smaller than 0.1\,mag. The uncertainty given in \citet{2011ApJ...737..103S} is below 0.01\,mag, but we assume the uncertainty of the $E(B-V)$ values to be at the level of 0.02\,mag. To calculate the total extinction in each passband, we used the \citet{1989ApJ...345..245C} and \citet{1994ApJ...422..158O} reddening law assuming $R_V$=3.1. The obtained values for passbands used in our study are $R_B$=4.133, $R_V$=3.136, $R_J$=0.892, $R_H$=0.553, $R_K$=0.363, $R_g$=3.713, $R_r$=2.693, $R_i$=2.108. 

\begin{table} 
    \caption{New optical and near$-$infrared photometry for Anomalous Cepheids analyzed in this paper. The full version of this Table is available as supplementary material in a machine readable format.}
    \label{tab:photometry}
    \centering
    \begin{tabular}{ccccc}
    \hline\hline
   
    $Star$ & $Filter$ & $HJD$ & $m$ & $\sigma _{m}$\\
    & & (days) & (mag) & (mag) \\
    \hline
    XZ Cet &	V	& 2458342.531808 & 9.792 & 0.005\\
XZ Cet	&	V	&	2458344.533328	&	10.386	&	0.005\\
XZ Cet	&	V	&	2458346.538110	&	9.894	&	0.005\\
XZ Cet	&	V	&	2458347.533630	&	10.441	&	0.005\\
XZ Cet	&	V	&	2458348.547888	&	10.470	&	0.005\\
XZ Cet	&	V	&	2458349.540027	&	10.226	&	0.005\\
... & ... & ... & ... & ...\\
    \hline
    \end{tabular}
    \end{table}

    

        \begin{table*}
            \caption{Data of the Milky Way Anomalous Cepheids.}
            \label{tab:acep_data_cd}

            \centering
            \begin{tabular}{cccccccccc}
            \hline\hline
            $Name$ & $P$ & $\omega$ & $ZPO$ &  $RUWE$ & $GOF$ & $G$ & $Bp-Rp$ & $d_g$ & $E(B-V)$\\
            & (day) & (mas) &  (mas) & & & (mag) & (mag) & (pc) & (mag) \\
            (1)& (2) & (3) &  (4) & (5)& (6)& (7) & (8) & (9) & (10) \\
            \hline
            V338 Pup$^a)$ & 0.79732 & 1.7861$\pm$0.0138 & $-$0.0086 & 1.06 & 1.26  & 9.05 & 0.65 & 557$^{+4}_{-4}$ & 0.042 \\
            XZ Cet & 0.82316 & 0.8540$\pm$0.0188 & $-$0.0262 & 0.86 & $-$3.23  & 9.37 & 0.59 & 1142$^{+24}_{-27}$ & 0.020 \\
            FY Aqr & 1.02289 & 0.2938$\pm$0.0198 & $-$0.0259 & 1.22 & 3.79 & 12.39 & 0.78 & 3138$^{+160}_{-166}$ & 0.075\\
            V716 Oph & 1.11592 & 0.4027$\pm$0.0228 & $-$0.0262 & 1.45 & 12.34  & 12.04 & 1.13 & 2339$^{+110}_{-136}$ & 0.375 \\
            BF Ser & 1.16545 & 0.2070$\pm$0.0217 & $-$0.0299 & 1.70 & 16.53  & 12.11 & 0.56 & 4275$^{+370}_{-364}$ & 0.028 \\
            CE Her & 1.20945 & 0.2529$\pm$0.0191 & $-$0.0130 & 1.56 & 13.65  & 12.45 & 0.80 & 3604$^{+225}_{-218}$ & 0.068 \\
            XX Vir & 1.34820 & 0.1440$\pm$0.0194 & $-$0.0277 & 1.34 & 6.24  & 12.27 & 0.73 & 5847$^{+866}_{-641}$ & 0.034\\
            UY Eri & 2.21331 & 0.2752$\pm$0.0215 & $-$0.0356 & 0.96 & $-$0.72  & 11.24 & 0.74 & 3215$^{+230}_{-200}$ & 0.058 \\
            \hline
            \end{tabular}
            \tablefoot{Table contains pulsational periods (column 2), parallaxes (column 3), $Gaia$ parallax zero$-$point offset (column 4), $Gaia$ parallax Renormalised Unit Weight Error (column 5), $Gaia$ parallax Goodness$-$of$-$Fit (column 6), $Gaia$ $G$ passband magnitude (column 7), $Gaia$ $Bp-Rp$ color index (column 8), geometric distance from \citet{2021AJ....161..147B} (column 9), and reddening (column 10).
            $^{a)}$ Star classified as RR Lyrae type.}
            \end{table*}

            \begin{table*}
                \caption{Mean magnitudes of the Milky Way Anomalous Cepheids in the Johnson $B$, $V$, 2MASS $J$, $H$, $K_{\mathrm{S}}$, and Pan$-$STARRS $g$, $r$, $i$  passbands.}
                \label{tab:acep_data}
                \centering
                \begin{tabular}{ccccccccc}
        
                \hline\hline
                $Name$ & $<B>$ &  $<V>$ &  $<J>$ &  $<H>$ & $<K_{s}>$ & $<g>$ &  $<r>$ &  $<i>$\\
                 & (mag) & (mag) & (mag) &  (mag) & (mag)& (mag) &  (mag) & (mag)\\
                \hline
                V338 Pup & $-$ & $-$ &$-$ &$-$ &$-$ &9.33$\pm$0.01 & 9.08$\pm$0.01 & 8.99$\pm$0.01 \\
                XZ Cet & 9.83$\pm$0.01 & 9.50$\pm$0.01  &  8.55$\pm$0.01 & 8.33$\pm$0.01 & 8.25$\pm$0.02& 9.65$\pm$0.01 & 9.44$\pm$0.01 & 9.41$\pm$0.01\\
                FY Aqr & $-$ & $-$ &$-$ &$-$ &$-$ & 12.56$\pm$0.01 & 12.28$\pm$0.01 & 12.17$\pm$0.01 \\
                V716 Oph & 12.77$\pm$0.01 & 12.14$\pm$0.01  & 10.43$\pm$0.01 & 10.11$\pm$0.01 & 9.98$\pm$0.01 & $-$ & $-$ & $-$ \\
                BF Ser & 12.37$\pm$0.01 & 12.07$\pm$0.01  & 11.23$\pm$0.01 & 10.00$\pm$0.01 & 10.94$\pm$0.02 & $-$ & $-$ & $-$\\
                CE Her & 12.64$\pm$0.02 & 12.26$\pm$0.02  & 11.45$\pm$0.02 & 11.16$\pm$0.02 & 11.09$\pm$0.02& $-$ & $-$ & $-$\\
                XX Vir & 12.69$\pm$0.01 & 12.33$\pm$0.01 & 11.33$\pm$0.01 & 11.08$\pm$0.01 & 11.04$\pm$0.01& 12.43$\pm$0.01 & 12.24$\pm$0.01 & 12.18$\pm$0.01\\
                UY Eri & 11.79$\pm$0.01 & 11.34$\pm$0.01 & 10.12$\pm$0.01 & 9.85$\pm$0.01 & 9.75$\pm$0.01& 11.49$\pm$0.01 & 11.21$\pm$0.01 & 11.09$\pm$0.01\\
                \hline

                \end{tabular}
                \end{table*}

\subsection{Spectroscopy}

For three ACeps from our MW sample we collected time$-$series of spectra with four high resolution (40,000$-$100,000) spectrographs hosted by the European Southern Observatory (ESO): CORALIE installed on the 1.2m ``Swiss'' Leonhard Euler telescope at La Silla Observatory \citep{2000AA...354...99Q}; HARPS installed on the 3.6m telescope at La Silla Observatory \citep{2000SPIE.4008..582P}; FEROS installed on the 2.2m MPG/ESO telescope at La Silla Observatory \citep{1998SPIE.3355..844K}, and UVES installed on the 8.2m VLT UT2 (Kueyen) telescope at Paranal Observatory \citep{2000SPIE.4005..121D}. 

Spectra from HARPS and UVES were reduced (instrumental calibrations and wavelength solution) using dedicated pipelines offered by ESO. CORALIE and FEROS spectra were reduced with the \texttt{CERES} pipeline \citep{2017PASP..129c4002B}. If needed, we merged the echelle orders with a custom Python script developed by our team and normalized spectra by modelling the continuum with polynomials (after excluding the most prominent lines). We restricted the analysis to the wavelength range between 450nm and 650nm and we used a mask to exclude telluric lines and problematic regions (like gaps) present in some spectra.

Radial velocities (RVs) were measured using the Broadening Function (BF) technique \citep{2002AJ....124.1746R} implemented in \texttt{Ravespan} \citep{2017ApJ...842..110P}. We used the Gaussian profile to model the BF, however, it is important to note that the projection factor depends quite strongly on the function used to model the BF (or cross$-$correlation function profile) as shown by e.g. \citet{2023AA...671A..14N} and \citet{2024AA...689A.241W}. We selected the Gaussian profile to be consistent with \citet{2021AA...656A.102T}, \citet{2024AA...684A.126B}, \citet{2024AA...689A.241W}, \citet{2024AA...690A.295Z}. A typical uncertainty of individual RVs measurements is 0.1$-$0.15 km/s. We do not observe any obvious systematic shifts between the measurements from different instruments. Our RV measurements are presented in Table \ref{tab:rvs}.

\begin{table} 
    \caption{The new radial velocity measurements for the three Anomalous Cepheids analyzed in this paper. Full version of this Table is available as supplementary material in a machine readable format.}
    \label{tab:rvs}
    \centering
    \begin{tabular}{ccccc}
    \hline\hline
   
    $Star$ & $HJD$ & $RV$ & $\sigma _{RV}$ & $instrument$\\
    & (days) & (km/s) & (km/s) & \\
    \hline
    V716 Oph & 2457864.81823  & $-$334.39 & 0.16 & CORALIE\\
    V716 Oph & 2457865.82941 & $-$332.38 &  0.16 & CORALIE\\
    V716 Oph & 2457866.84505 & $-$337.70 &  0.13 & CORALIE\\
    V716 Oph & 2457867.86937 & $-$342.46 &  0.08 & CORALIE\\
    V716 Oph & 2457868.81033 & $-$355.67 &  0.09 & CORALIE\\
    V716 Oph & 2458233.88310 & $-$342.25 &  0.10 & HARPS\\
    V716 Oph & 2458247.86930 & $-$383.37 &  0.58 & HARPS\\
    V716 Oph & 2458559.77147 & $-$340.57 &  0.09 & HARPS\\
    V716 Oph & 2458559.90268 & $-$330.82 &  0.13 & HARPS\\
    ... & ... & ... & ... & ...\\
    \hline
    \end{tabular}
    \end{table}

\subsection{Distances}

Distances of the analyzed stars are derived from parallaxes from the $Gaia$ Data Release 3 \citep[DR3,][]{2021A&A...649A...2L}. The values of parallaxes and uncertainties are given in Table~\ref{tab:acep_data_cd}.  Post$-$processing of $Gaia$ DR3 catalogue by \citet{2021A&A...649A...4L}, assuming distant quasars as a "stable" reference frame, resulted in the estimation of the parallax zero$-$point correction for different locations in the sky. Moreover, the authors used binary stars and Large Magellanic Cloud stars to measure the magnitude and color dependent corrections to systematic shifts of parallaxes which are also present due to chromatic effects in the $Gaia$ optical system. We calculated zero$-$point offsets resulting from this study with the dedicated \texttt{Python} code described in the same work and they are presented in Table \ref{tab:acep_data_cd}. These corrections are subtracted from parallaxes in the analysis. The size of the zero$-$point corrections for $Gaia$ parallaxes is however still debated in the literature \citep[e.g.][]{2021AA...654A..20G}. Using $Gaia$ DR3 parallaxes and the model of the MW as a prior in the Bayesian approach, \citet{2021AJ....161..147B} estimated the distances of nearly 1.5 billion stars. Distances of our sample stars resulting from this work are presented in column 9 ($d_g$) of Table \ref{tab:acep_data_cd}.

\begin{figure*}[]
    \centering
    \includegraphics[width=0.8\textwidth]{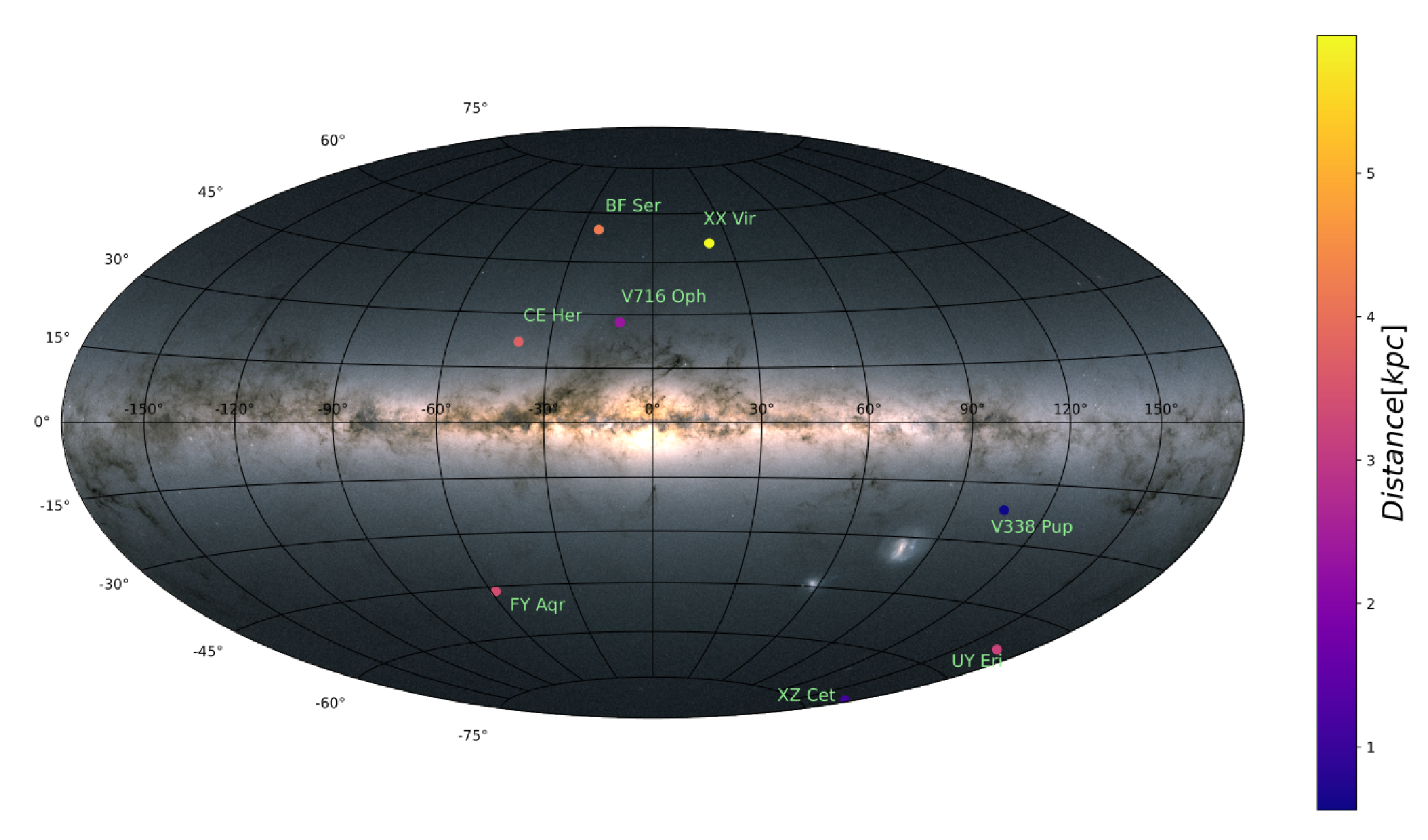}
    \caption{The Gaia photometric map of the Milky Way with marked positions of Anomalous Cepheids considered in this work. Color denotes the distance of a given star from the Sun\label{fig:mw_acep_map}}
    \end{figure*}

After applying zero$-$point offset, parallaxes range between 1.78 and 0.17mas (corresponding to distances $\sim$560$\,$pc and $\sim$5900$\,$pc) and relative uncertainties of parallaxes between 0.8\% and 20\% with a median value of 6.7\%. In case of the most distant stars the zero$-$point corrections are also quite significant (~20\%). We decided to not discard the stars based on parallax relative errors. For three stars in the sample, the Renormalised Unit Weighted Error (RUWE) and the Goodness Of Fit (GOF) parallax quality parameters given in the $Gaia$ catalogue and presented in Table \ref{tab:acep_data_cd} exceed the limits for good quality parallaxes \citep[$RUWE$<1.4, $GOF$<12.5,][]{2021A&A...649A...2L}, which indicates additional movement of the photocenter due to e.g. binarity. Given the low number of stars in the sample we do not reject them from the analysis based on these parameters. We expect that parallaxes of these stars will be improved in the next $Gaia$ Data Release. Figure \ref{fig:mw_acep_map} shows the map of the MW with marked positions of the analysed ACeps and their distances represented by colors.

\section{Classification}
\begin{figure}[]
    \centering
    \includegraphics[width=0.5\textwidth]{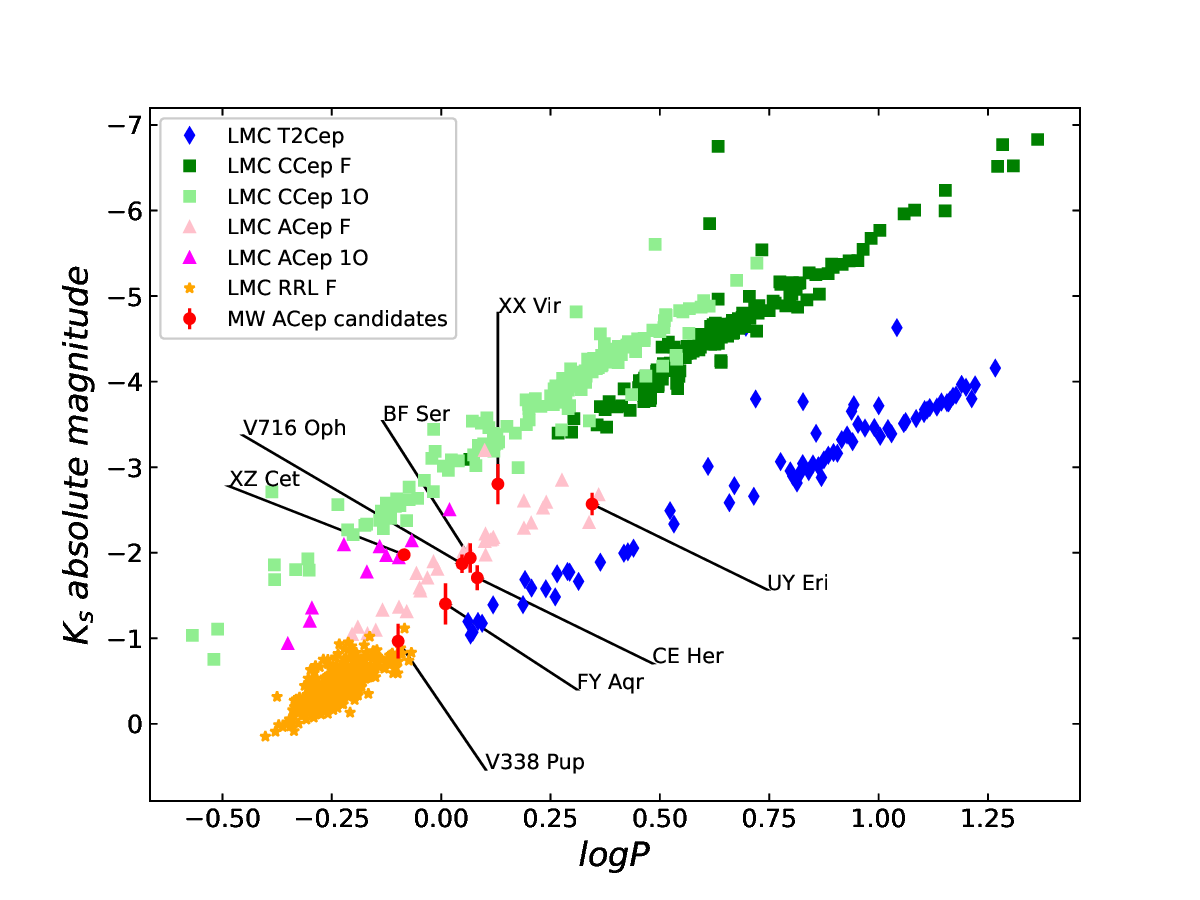}
    \caption{Period$-$luminosity diagram for nearby Anomalous Cepheids candidates and LMC Classical (fundamental mode and first overtone), Anomalous (fundamental mode and first overtone) and Type II Cepheids and RR Lyrae (fundamental mode) type stars.\label{fig:pl_classification}}
    \end{figure}

Figure \ref{fig:pl_classification} shows the period$-$luminosity diagram in the $K_{\mathrm{S}}$ passband for MW Anomalous Cepheids and LMC Anomalous (fundamental mode and first overtone), Classical (fundamental mode and first overtone) and Type II Cepheids and RR Lyrae (fundamental mode) stars. For V338~Pup and FY~Aqr we are missing the near$-$infrared light curves from IRIS so we adopted the measurement from the 2MASS catalogue and an uncertainty of 0.2\,mag, which is the expected semi$-$amplitude of the $K_{\mathrm{S}}$ light curve. Absolute magnitudes of the MW sample are calculated using distance moduli obtained from parallaxes:
 \begin{equation}\label{eq:dm}
    (m-M)_0=-5\log \omega +10.
    \end{equation}
Photometry of LMC stars comes from the VISTA Magellanic Clouds (VMC) survey \citep{2012MNRAS.424.1807R,2021MNRAS.504....1C,2024A&A...685A..41S,2025AA...699A.370S} and was corrected for extinction using $E(B-V)$ values from the \citet{2020ApJ...889..179G} reddening maps. The distance modulus of the LMC from \citet{2019Natur.567..200P} was subtracted to obtain absolute magnitudes. We show this plot to discuss the classification of our MW stars as ACeps.

BF~Ser and V716~Oph were traditionally considered Type II Cepheids \citep[e.g.][]{2018MNRAS.477.2276K}, but their ACep nature was first suspected by \citet{2014ApJS..213....9D} and subsequently confirmed by \citet{2018SerAJ.197...13J} and \citet{2023A&A...674A..17R}. Based on the position of V716~Oph and BF~Ser on the period$-$luminosity diagram we conclude, that they are indeed ACeps.

CE~Her, FY~Aqr and UY~Eri were classified as ACeps by \citet{2018SerAJ.197...13J}. Classification of CE~Her as an ACep was also confirmed by \citet{2023A&A...674A..17R}. Based on the position of CE~Her and UY~Her on the period-luminosity diagram in Figure \ref{fig:pl_classification}, we also classify these stars as ACeps. FY~Aqr is located between LMC ACeps and short period Type II Cepheids (BL Her type stars) in our period-luminosity diagram, which is most probbly the result of the high uncertainty of the mean magnitude as it comes from single 2MASS measurement. The shape of the $V$ passband light curve indicates that it is most likely an ACep so we include this star in our analysis.

XZ~Cet is known multi$-$mode Anomalous Cepheid \citep{2021ApJS..253...11P} showing amplitude and period variations in light curve and radial velocity curve \citep{2007A&A...461..613S}. On our peirod$-$luminosity diagram it is located among the LMC overtone ACeps, therefore we also confirm the classification of this star.

V338~Pup is the nearest star in our sample. In the AAVSO database it is classified as an ACep, while according to the SIMBAD database and in \citet{2022MNRAS.513..788G} it is the RR Lyrae type star. The location of this star in our period$-$luminosity diagram is not conclusive as the uncertainty of the mean magnitude is large and based on this it can be considered as short period fundamental mode ACep or long period RR Lyrae ab star. The shape of the optical light curve is not typical for ACeps thus it is more likely an RR Lyrae type star. We do not take into account this star during the fitting of PLRs.

The most distant star in our MW sample, XX~Vir, lies significantly above the PLR of LMC ACeps pulsating in fundamental mode, on the extension of the first overtone ACeps towards longer periods or fundamental mode Classical Cepheids PLR towards very short periods. The shape of the light curve of this star is typical for stars pulsating in the fundamental mode in the Instability Strip so it is very unlikely to be an overtone pulsator. Its location in the Milky Way far above the Galactic disc (see Figure \ref{fig:mw_acep_map}) and a very low metallicity of this star of about $-$1.5\,dex \citep{2018MNRAS.477.2276K} exclude the possibility that this star is a Classical Cepheid. \citet{2018SerAJ.197...13J} classified this star as an ACep. XX~Vir was often considered as Type II Cepheids but the shape of its light curve is not typical for this type of pulsators and it significantly deviates from the PLR of Type II Cepheids in our Figure \ref{fig:pl_classification}. We suspect that the obtained high absolute brightness of this star, being indeed an ACep, is the result of the underestimated parallax in the $Gaia$ DR3. This conclusion is also supported by the BW analysis presented in Section \ref{sec:analysis_bw}. We decided to not remove this star from the final sample as we perform a weighted fit and given the high uncertainty of the absolute magnitude resulting from the big relative uncertainty of the parallax it does not influence our final results for PLRs significantly$-$the obtained zero$-$points are smaller by about 0.02\,mag when we exclude XX~Vir.

\section{Methodology, results and discussion}\label{sec:analysis}

\subsection{Period$-$Luminosity Relations}

\begin{figure}[]
            \centering
            \includegraphics[width=0.45\textwidth]{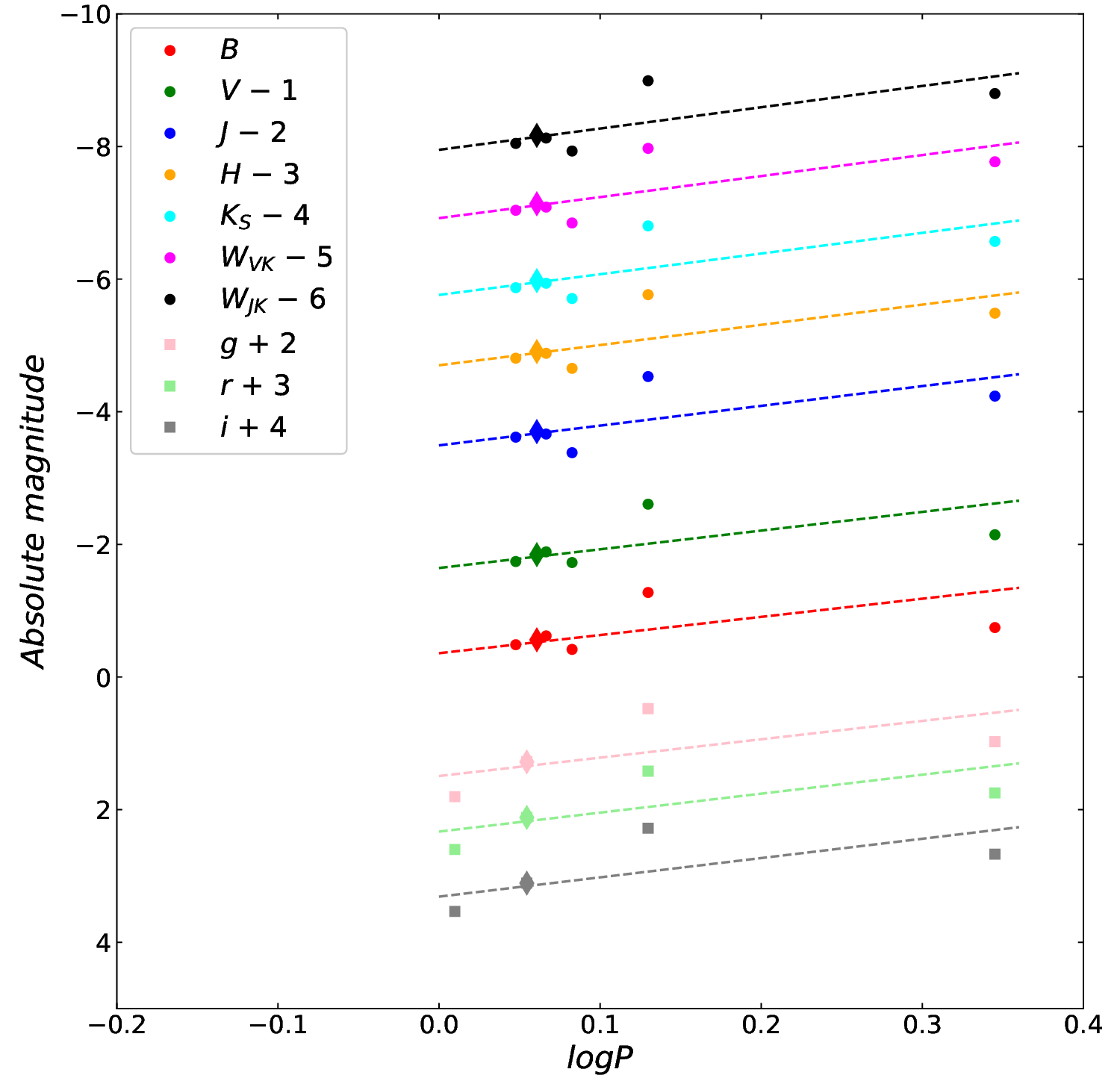}
            \caption{Period$-$Luminosity relations for MW Anomalous Cepheids in the Johnson $B$, $V$, 2MASS $J$, $H$, $K_{\mathrm{S}}$, and Pan$-$STARRS $g$, $r$, $i$  passbands and $W_{VK}$ and $W_{JK}$ Wesenheit indices. A first overtone star $-$ XZ~Cet $-$ with fundamentalized period is plotted with a diamond symbol.\label{fig:mw_acep_plr}}
            \end{figure}

Figure \ref{fig:mw_acep_plr} shows the period$-$luminosity diagram for our MW ACeps in all the considered passbands and Wesenheit indices. The period of overtone ACep XZ~Cet is fundametalized by adding the value of 0.139 to the $\log P$. This value is a logarithm of the fundamental$-$to$-$overtone period ratio of OGLE$-$GAL$-$ACEP$-$091 multimode Anomalous Cepheid, adopted from \citet{2020ApJ...901L..25S}. This value is slightly lower than obtained by \citet{2025AA...699A.370S} from minimizing the scatter of common PLRs for fundamental and overtone ACeps in the LMC, which amounts to 0.145. Later we discuss the influence of the adopted shift to fundamentalize the period of XZ~Cet on our results.

Wesenheit indices are defined as:
\begin{equation}
W_{XY} = Y - \frac{A_Y}{E(X-Y)}(X-Y)
\end{equation}
where $X$ and $Y$ are magnitudes in two selected filters and $A_Y/E(X-Y)$ is the total$-$to$-$selective extinction ratio, which can be calculated assuming a reddening law. We consider two Wesenheit indices: $W_{VK}=K_\mathrm{S}-0.14(V-K_\mathrm{S})$ and $W_{JK}=K_\mathrm{S}-0.69(J-K_\mathrm{S})$, to be consistent with \citet{2025AA...699A.370S}.

We want to obtain the slope $a$ and the zero$-$point $b$ of the PLR in form: 
\begin{equation}\label{eq:plr}
M = a (\log P -0.1) +b,
\end{equation}
where the numerical value $0.1$ corresponds to the pivot period adopted for the sample. As the number of stars in our sample is very low, we decided to adopt the slopes of PLRs determined by \citet{2025AA...699A.370S} from LMC ACeps for the $V$, $J$ and $K_\mathrm{S}$ passbands and the Wesenheit indices. Photometry of the LMC ACeps in the $B$, $H$, $g$, $r$ and $i$ passbands is not available. As shown by \citet{2025AA...699A.370S}, the slopes of PLRs of ACeps depend linearly on the inverse of the effective wavelength. In the case of fundamental mode ACeps they obtained the slope of 0.188$\pm$0.12\,mag/dex/nm$^{-1}$ for the wavelength dependence of $a$. We used this value and the $V$ as a reference filter to find the slopes in $B$, $H$, $g$, $r$ and $i$; the corresponding values are $-$2.74, $-$3.05, $-$2.77, $-$2.86, and $-$2.91, respectively. In the case of the Small Magellanic Clouds Anomalous Cepheids, the authors found a stronger dependence of 0.321$\pm$0.16\,mag/dex/nm$^{-1}$ and the slope of the PLR in the $V$ passband of $-$2.43$\pm$0.10, which is lower by ~0.3 than for LMC ACeps. The slopes for the $J$ and $K_\mathrm{S}$ and Wesenheit indices are lower by ~0.2 for SMC ACeps. Adopting the SMC $V$ passband slope and dependence of slopes on the wavelength we obtained the slopes of $-$2.29, $-$2.82, $-$2.34, $-$2.50 and $-$2.58 for $B$, $H$, $g$, $r$ and $i$, respectively. The difference between the LMC and SMC slopes is ~0.4 for the bluest filters and moving to the infrared it decrease to ~0.2. At this point it is not clear if different slopes for these two populations of ACeps can be related to the different mean metallicity as metallicity measurements for either LMC and SMC ACeps are not available in the literature. We assume that the similar difference of the slopes is possible for the MW sample but since we determine our zero$-$points for the pivot period ($\log P=$0.1), the impact of the assumed slopes is not significant. Changing the slope to the SMC value shifts the zero$-$point by ~0.02\,mag for either optical and infrared passbands. We conservatively add this value to the systematic error of the zero$-$points.

To minimize possible biases related to calculating distances from parallaxes \citep{1999ASPC..167...13A}, we use a Monte$-$Carlo simulation in which we draw a parallax $\omega$ of every star from the normal distributions defined by the original values of the parallaxes and their uncertainties. Then we calculate the distance moduli (Equation \ref{eq:dm}) which are subtracted from the magnitude $m$ of a given star in a given passband (corrected for extinction) to obtain the absolute magnitude $M$. Uncertainties of magnitudes and parallaxes are propagated, and we fit the line in the form given by Equation \ref{eq:plr} with the slope fixed to the LMC value using weighted least squares method. As a result we obtain the zero$-$point. We repeat this process 10 000 times and we obtain the histogram for the zero$-$point. The final value of the zero$-$point and its uncertainty are obtained by fitting the Gaussian profile to the histogram. These values are presented in Column 3 of Table \ref{tab:plr_mw_results}.

We also present the results obtained using $Astrometric$ $Based$ $Luminosity$ (ABL) as proposed by \citet{1997MNRAS.286L...1F} and \citet{1999ASPC..167...13A}:
\begin{equation}
    ABL = \omega 10^{0.2m-2}=10^{M/5}
    \end{equation}
Its advantage is that parallaxes are used directly and it is considered ''asymptotically unbiased'' when stars are not excluded based on the relative parallax error. The procedure of fitting ABL is also done using a Monte Carlo process. In each iteration we generated a random sample of magnitudes and parallaxes from the normal distributions defined by the mean magnitudes and parallaxes and their respective uncertainties. We fix the slope of the PLR and find the zero$-$point using the \texttt{curvefit} routine from the \texttt{Python SciPy} package. We repeat this process 10 000 times and we find the final zero$-$point and its uncertainty by fitting the Gaussian profile to the resulting histogram. The obtained zero$-$points are presented in Column 5 of Table \ref{tab:plr_mw_results}. They are in a very good agreement with zero$-$points obtained using the direct inverse of parallax  approach.

Finally, we used the distances from \citet{2021AJ....161..147B} to calculate the distance moduli:
\begin{equation}
(m-M)_0=5\log d - 5
\end{equation}
and subtracted them from apparent mean magnitudes to obtain absolute magnitudes. We propagated errors of distances and added them quadratically to the errors of mean magnitudes. We then used the weighted least$-$squares method to fit the PLR with the slope again fixed to the respective LMC value to obtain the zero$-$points. Uncertainties were again evaluated in 10 000 Monte Carlo simulations. The obtained zero$-$points are presented in Column 7 of Table \ref{tab:plr_mw_results}. They are significantly lower (by 0.05$-$0.06 mag) than zero$-$points obtained using parallaxes directly. 

\begin{figure}[]
            \centering
            \includegraphics[width=0.5\textwidth]{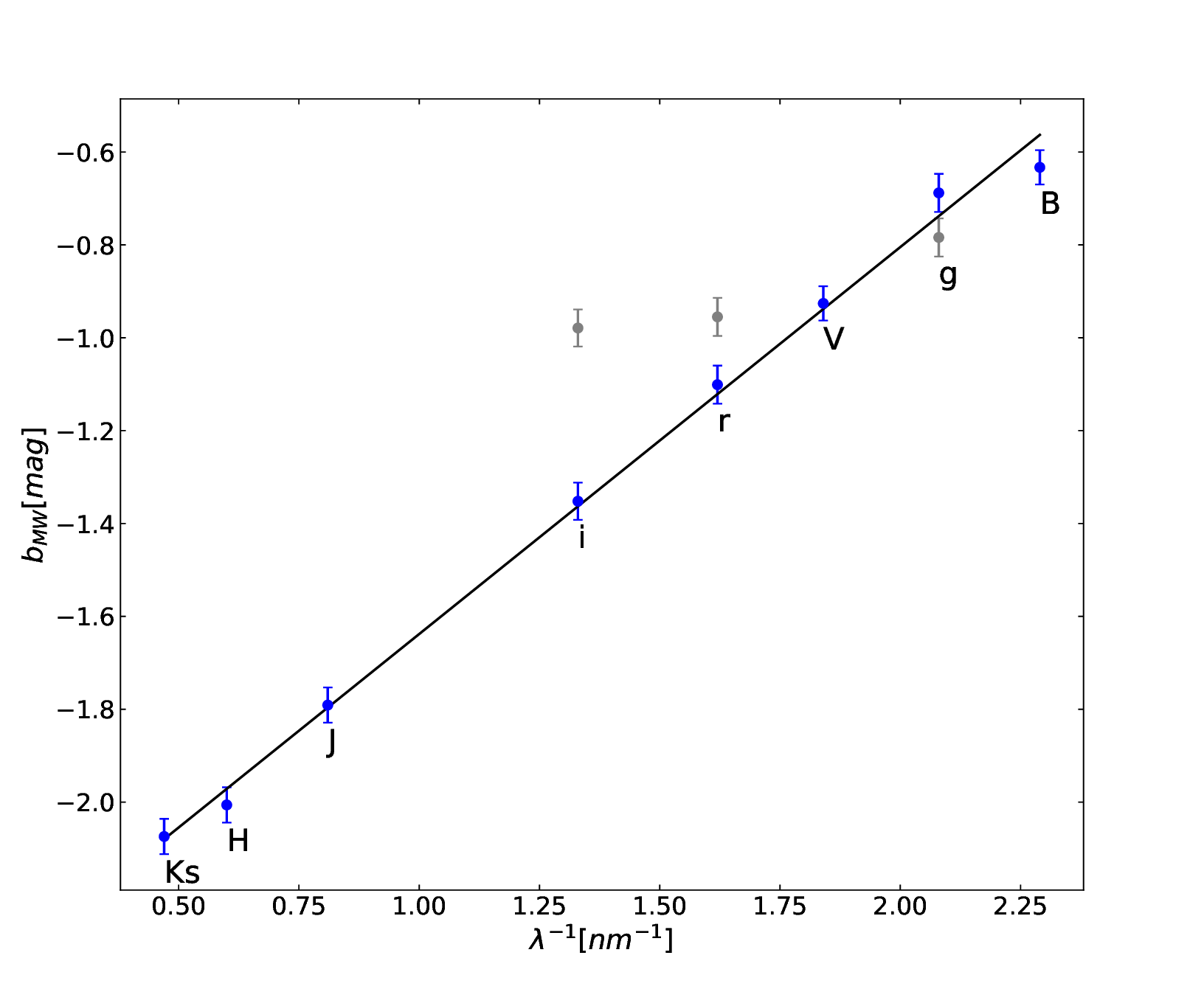}
            \caption{Zero$-$points of Period$-$Luminosity relations for Milky Way Anomalous Cepheids versus the inverse of the effective wavelength of the respective passband. For the $g$, $r$ and $i$ passbands two zero$-$points are given as originally the Pan$-$STARRS photometric system is in the AB system (grey points). We transformed the zero$-$points for these three passbands to the Vega system (blue points) using filter definitions from the VOSA database (see text for details). The black line is the least square fit for the zero$-$points in the Vega system.\label{fig:mw_acep_plr_zp}}
            \end{figure}

Figure \ref{fig:mw_acep_plr_zp} shows the zero$-$points of the PLRs versus the inverse of the effective wavelength of a given passband. As Pan$-$STARRS is in the AB system, while Johnson is in Vega system, we used filter definitions from VOSA \footnote{\url{https://svo2.cab.inta-csic.es/theory/vosa/}} to calculate the zero$-$point corrections for $g$, $r$ and $i$ to transform from AB to Vega system. We obtained the following corrections: 0.096, $-$0.146 and $-$0.373 for $g$, $r$ and $i$, respectively, which are added to the zero$-$points of the PLRs. In Figure \ref{fig:mw_acep_plr_zp} zero$-$points in the Vega system are plotted as blue points, while those in the AB system are shown in grey. We fitted the line to the relation formed by zero$-$points in the Vega system, and obtained a slope of 0.833$\pm$0.022mag/nm$^{-1}$, in perfect agreement with the values found by \citet{2025AA...699A.370S} for Large and Small Magellanic Clouds ACeps which are 0.840$\pm$0.056mag/nm$^{-1}$ and 0.856$\pm$0.048mag/nm$^{-1}$, respectively.

To estimate the systematic uncertainty of our zero$-$points we should take into account the systematic uncertainties of parallaxes, photometric zero$-$point, extinction, and assumed slopes. For the uncertainty of the photometric zero$-$point we assume 0.02\,mag for optical and 0.025\,mag for near$-$infrared passbands \citep[following ][]{2022ApJ...927...89W, 2024AA...689A.241W}. The systematic uncertainty of extinction is assumed to be 0.02\,mag. As described above, the uncertainty related to the adopted slopes is 0.02mag. The systematic uncertainty of parallaxes is related to the zero$-$point offset which is of the order of 0.02\,mas. We repeated the PLRs zero$-$point determination by shifting all parallaxes by such value and obtained a difference of the PLRs zero$-$point of about 0.09\,mag. Adding quadratically those three errors we obtain the total systematic uncertainty of 0.097\,mag and 0.098\,mag for optical and near$-$infrared passbands, respectively. 

Another source of uncertainty is the value added to fundamentalize the period of XZ~Cet star. In the presented analysis we adopted the value of 0.139 from \citet{2020ApJ...901L..25S}. If we adpot the value of 0.145 from \citet{2025AA...699A.370S}, the obtained zero$-$points differ by about 0.01\,mag. We add this value quadratically to the systematic errors and finally we obtain 0.099\,mag and 0.098\,mag for the optical and near$-$infrared passbands, respectively.

In Columns 4, 6 and 8 of Table \ref{tab:plr_mw_results} we present the distance modulus of the LMC calculated as the difference between zero$-$points of the LMC \citep{2025AA...699A.370S} and MW PLRs for a given band. The LMC distance modulus obtained directly from parallaxes using both linear fit and ABL method in each passband is in a very good agreement with the accurate value obtained in \citet{2019Natur.567..200P} from eclipsing binaries which amounts to 18.477$\pm$0.004(stat.)$\pm$0.026(syst.)\,mag. In contrast, using the distances from \citet{2021AJ....161..147B} gives higher LMC distance moduli than the canonical eclipsing binary value, but in agreement at 2$\sigma$ level.

\citet{2025AA...699A.370S} measured the LMC distance modulus using PLRs of ACeps in the $W_{G}$ and $W_{JK}$ Wesenheit indices. They used a sample of MW ACeps with $Gaia$ and 2MASS photometry and $Gaia$ DR3 parallaxes. Assuming the parallax zero$-$point offset to be $-$0.022\,mas they obtained the LMC distance modulus of 18.486$\pm$0.022 and 18.456$\pm$0.014 for $W_{G}$ and $W_{JK}$, respectively, with both values in very good agreement with our results. It is, however, important to emphasize that the influence of metallicity on the ACeps absolute magnitudes and the metallicity of LMC ACeps are still an open question and the metallicity effect is not taken into account in either study.

\citet{2004AA...417.1101M} used theoretical ACeps models to derive period$-$luminosity and period$-$luminosity$-$color relations (PLCRs). They presented the mass-dependent PLCRs for the $V$ passband and the $W_{VK}$ index (see their Tables 2 and 4):
\begin{equation}
\langle M_V \rangle = -1.99-3.05\log P + 1.31(V-K) -1.86 \log M,
\end{equation}

\begin{equation}
\langle W_{VK} \rangle = -1.73-2.93\log P -1.83 \log M.
\end{equation}
 The mean $(V-K_{\mathrm{S}})$ color index of our ACeps amounts to 1.16. To reproduce our zero$-$points we have to use $M=$1.8$M_{\odot}$ for $V$ passband, and $M=$2.85$M_{\odot}$ for Wesenheit index. While the first value is in the range of expected masses for Anomalous Cepheids, the second is a bit higher than the usually assumed upper limit for the mass which is 2.5$M_{\odot}$ \citep{2024AA...682A...1R}.

        \begin{table*}
            \centering
            \caption{Period$-$Luminosity relations of Anomalous Cepheids in the Milky Way and the derived distance modulus of the LMC (see text for details).}
            \label{tab:plr_mw_results}
            \begin{tabular}{cccccccc}
            \hline\hline
            
            Filter & $a$ &  $b_{MW}$ &  $\mu_{LMC}$& $b_{MW,ABL}$ &  $\mu_{LMC,ABL}$ & $b_{MW,BJ}$ &  $\mu_{LMC,BJ}$\\
             & & (mag) & (mag) & (mag) & (mag) & (mag) & (mag)\\
            (1) & (2) & (3) & (4) & (5) & (6) & (7) & (8)\\
            \hline
            $B$ & $-$2.74 & $-$0.633$\pm$0.037  & $-$& $-$0.649$\pm$0.037& $-$ & $-$0.712$\pm$0.050& $-$\\
            $V$ & $-$2.82 & $-$0.926$\pm$0.037  & 18.445$\pm$0.052& $-$0.939$\pm$0.037 &18.458$\pm$0.052 & $-$1.014$\pm$0.049&18.533$\pm$0.059\\
            $J$ & $-$2.98 & $-$1.791$\pm$0.038 & 18.455$\pm$0.047& $-$1.796$\pm$0.038& 18.460$\pm$0.047 & $-$1.867$\pm$0.051&18.531$\pm$0.056\\
            $H$ & $-$3.05 & $-$2.006$\pm$0.037  & $-$ & $-$2.011$\pm$0.038& $-$ & $-$2.091$\pm$0.049& $-$\\
            $Ks$ & $-$3.12 & $-$2.074$\pm$0.038  &18.448$\pm$0.045 & $-$2.080$\pm$0.039 & 18.454$\pm$0.046 &$-$2.155$\pm$0.049& 18.529$\pm$0.055\\
            $W_{VK}$ & $-$3.17 & $-$2.237$\pm$0.038  & 18.461$\pm$0.044& $-$2.242$\pm$0.039 & 18.466$\pm$0.045 & $-$2.317$\pm$0.049& 18.541$\pm$0.054\\
            $W_{JK}$ & $-$3.21 & $-$2.271$\pm$0.039  & 18.454$\pm$0.045 & $-$2.276$\pm$0.039 & 18.459$\pm$0.045 &$-$2.354$\pm$0.049 & 18.537$\pm$0.055\\
            $g$ & $-$2.77 & $-$0.784$\pm$0.041&$-$& $-$0.790$\pm$0.041&$-$ & $-$0.777$\pm$0.053& $-$\\
            $r$ & $-$2.86 & $-$0.955$\pm$0.041&$-$& $-$0.963$\pm$0.041& $-$ & $-$0.971$\pm$0.051& $-$\\
            $i$ & $-$2.91 & $-$0.979$\pm$0.040&$-$& $-$0.985$\pm$0.041& $-$ & $-$1.017$\pm$0.054& $-$\\
            \hline
            \end{tabular}
            \tablefoot{Period$-$luminosity relations have the form $M=a(\log{P-0.1})+b$.}
            \end{table*}

\subsection{Projection factor and radii}\label{sec:analysis_bw}
\begin{figure}[]
    
    \includegraphics[width=0.5\textwidth]{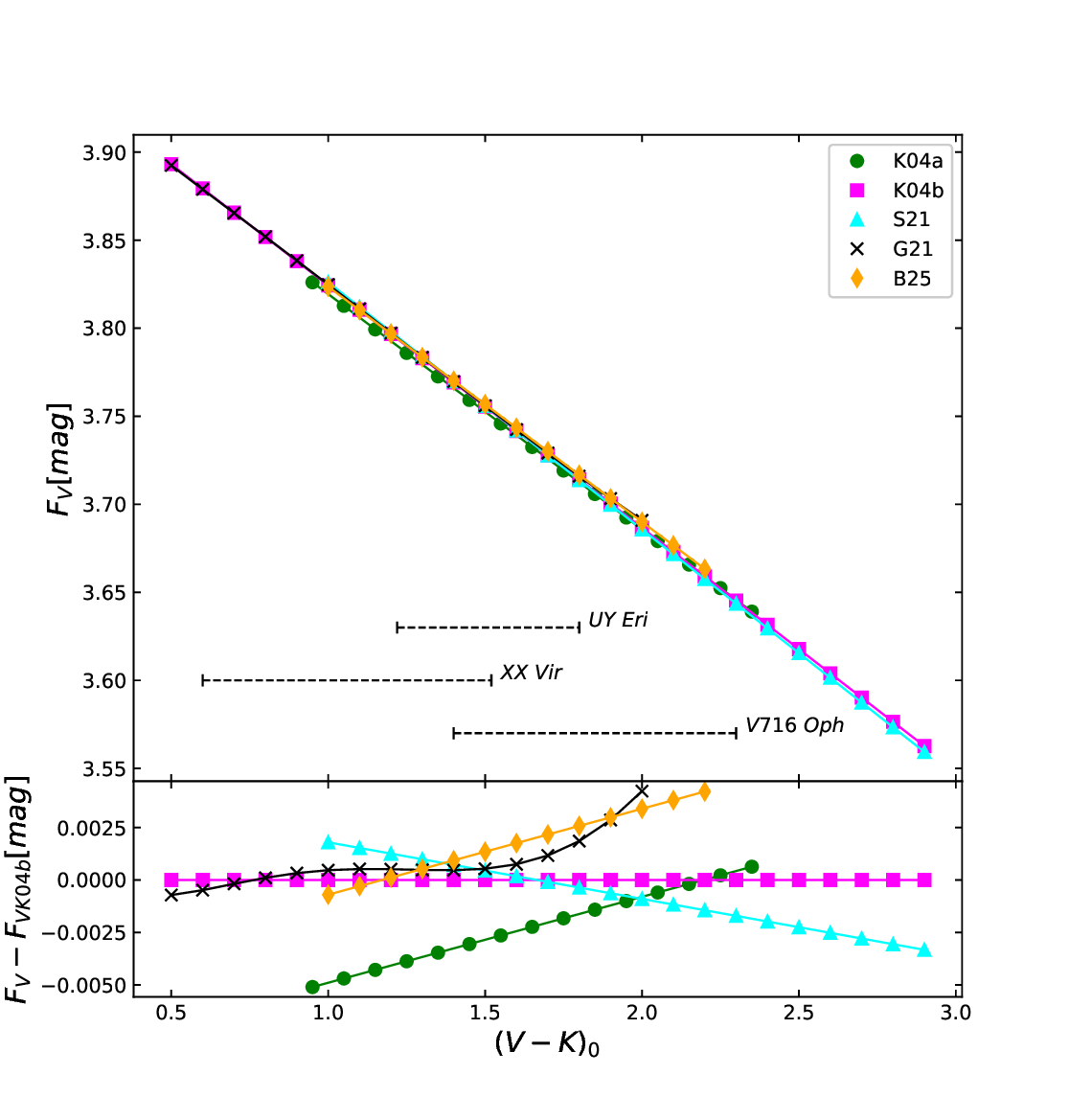}
    \caption{Surface Brightness-Color Relations used to calculate angular diameters of the three Anomalous Cepheids. Dashed section denotes the color index range of each analyzed star.\label{fig:irsb}}
    \end{figure}

Our implementation of the Baade$-$Wesselink method used in this study is described in detail in \citet{2024AA...689A.241W} and is based on the surface brightness techique as proposed by \citet{1976MNRAS.174..489B} and developed by \citet{1994AJ....108.1421W} and \citet{1997AA...320..799F} \citep[see also][and references therein]{2011AA...534A..94S,2018AA...620A..99G}. The apparent angular diameter $\theta$ of the star is calculated from the visual surface brightness defined as:


\begin{equation}\label{eq:sb}
    F_{V} := 4.2207-0.1V_0 - 0.5\log\theta = \log T_e +0.1BC,
\end{equation}
where $V_0$ is the extinction corrected visual magnitude, $T_e$ is the effective temperature, $BC$ is bolometric correction, and the numerical constant is the result of adopting the Sun as the zero$-$point. Effective temperature and bolometric correction are related to the color index, thus there is a relation between the surface brightness and the color index (SBCR) of the star for a given pulsational phase. There are several calibrations of such relations in the literature based on interferometric measurements of angular diameters of nearby stars. We use the calibration of $F_V$ versus the $(V-K)$ color index as it is very smooth compared to other color indices and almost insensitive to reddening \citep[see e.g. ][and references therein]{1997AA...320..799F,2004AA...428..587K,2021AA...649A.109G, 2024AA...690A.295Z}. Similarly as in \citet{2024AA...689A.241W} we use the relations from \citet[][hereafter K04a and K04b, respectively]{2004AA...428..587K,2004AA...426..297K}, \citet[][S21]{2021AA...652A..26S} and \citet[][G21]{2021AA...649A.109G} as they are independent, precise and cover the color range of the analyzed stars. Additionally, we perform the analysis using the very recent SBCR from \citet[][B25]{2025A&A...698A..46B}. Figure \ref{fig:irsb} presents the used SBCRs and the color range of each star. Using several independent relations allows us to estimate the uncertainty on the derived parameters resulting from the uncertainty of the surface brightness. 


    


In the BW method, variations of the angular size are compared to the physical radius displacement $\Delta R (\phi)$ in a corresponding pulsational phase $\phi$ to infer about the distance of the star. The physical radius displacement is calculated by integrating the radial velocity $V_r(\phi)$:
\begin{equation}\label{eq:integral}
\Delta R (\phi) = R(\phi) - \langle R \rangle = \int_{0}^{\phi} -p P \left (V_r(\phi ') - V_{\gamma} \right )d \phi '
\end{equation}
where $\langle R \rangle$ is the mean radius of the star, $p$ is the projection factor, $P$ is the pulsational period (expressed in seconds as velocity is in km/s), and $V_{\gamma}$ is the mean (systemic) radial velocity. The projection factor is needed to transform the radial velocity as measured from the Doppler shift of absorption lines in spectra to the velocity of the expanding or contracting stellar atmosphere (called pulsational velocity). Here we aim at determining this parameter for three Anomalous Cepheids from our sample.

The angular size for a given pulsational phase $\theta(\phi_i)$ and the physical radius $R(\phi_i)$ are connected by the following geometrical formula:
\begin{equation}\label{eq:bw_0}
    \begin{aligned}
    \theta (\phi_i)= 6.6746\times10^{-9} \times 2 \omega R(\phi_i)
    \end{aligned},
\end{equation}
where $\omega$ is the parallax (in milliarcseconds). The numerical constant comes from changing parsec to kilometres and radians to milliarcseconds; in fact it is simply the inverse of the astronomical unit. Using Equation \ref{eq:integral}, we obtained the following relation:
\begin{equation}\label{eq:bw}
    \begin{aligned}
    \theta (\phi_i)= 6.6746\times10^{-9}\times2\omega p \left (\int_{0}^{\phi} -P \left (V_r(\phi ') - V_{\gamma} \right )d \phi ' \right ) + \langle \theta \rangle
    \end{aligned},
\end{equation}
where $\langle \theta \rangle$ is the mean angular diameter of the star. By fitting a straight line to this relation, we obtain the projection factor (as the slope) and the mean angular diameter of the star (the zero$-$point), which is used to calculate the mean radius.

\begin{figure*}[]
    \centering
    \includegraphics[width=0.8\textwidth]{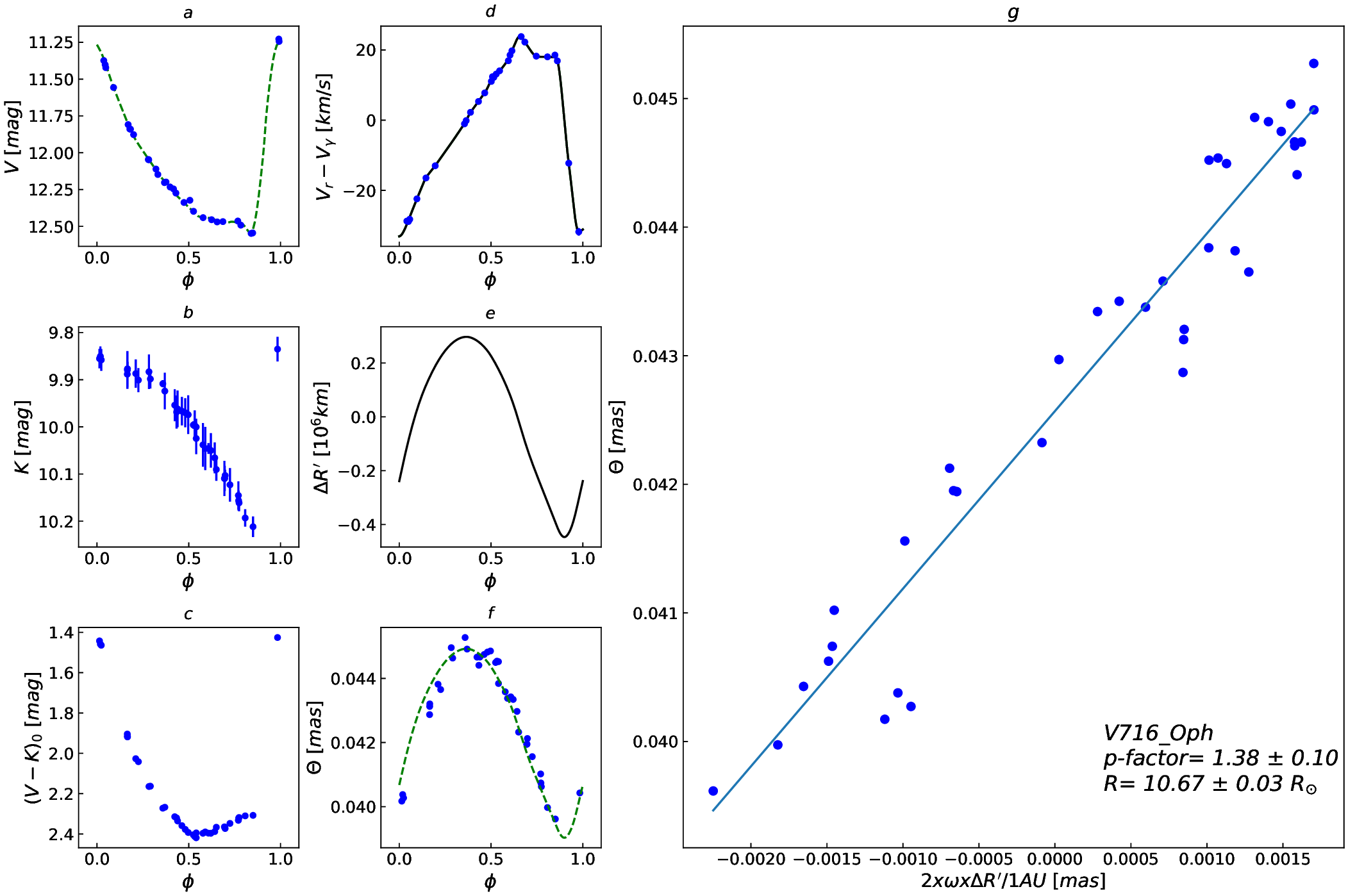}
    \caption{The Baade$-$Wesselink analysis of V716 Oph. Panels $a$, $b$, and $d$ show our $V$, $K_{\mathrm{S}}$, and radial velocity measurements, respectively. Panel $c$ shows the unreddened colour index curve. Green dashed line in panel $a$ is the template fitted to the measurements as described in the text. Black lines in these plots are $Akima$ splines fitted to the measurements. Panel $e$ shows the integrated radial velocity curve. Panel $f$ shows the angular diameter measured using the $K04b$ SBCR. The green line in this panel is the curve from panel $e$, rescaled using Equation \ref{eq:bw} and the measured values of the p$-$factor and the mean radius. Panel $g$ shows the relation between angular diameters from panel $f$ and the corresponding values of the integrated radial velocity from panel $e$, rescaled using Equation \ref{eq:bw}. The slope of this relation is the p$-$factor, and the zero$-$point is the mean angular diameter used to calculate the mean radius of the star.\label{fig:v716_oph_bw}}
    \end{figure*}

Figures \ref{fig:v716_oph_bw}, \ref{fig:xx_vir_bw} and \ref{fig:uy_eri_bw} show the phased light curves (panels $a$, $b$) and radial velocity curves (panel $d$)  corrected for period changes as described in Section \ref{sec:data_mw}, color index curve (panel $c$), integrated radial velocity curve (panel $e$) and angular diameter over the pulsational cycle (panel $f$). In our approach, it is required to properly trace the course of the $V$ passband light curve. In case of V716~Oph and XX~Vir our VYSOS16 photometry is sparse, so we used the All Sky Automated Survey for Supernovae (ASAS$-$SN) \citep{2017PASP..129j4502K} and the All Sky Automated Survey (ASAS) \citep{2002AcA....52..397P} $V$ passband photometry, respectively, to construct templates for our $V$ passband light curves. We created the templates by fitting $Akima$ splines to the literature photometry of each star. The template (namely the phase shift, mean magnitude and amplitude) was then fitted to our VYSOS16 photometry using $\chi^2$ minimisation.

Panels $g$ of Figures \ref{fig:v716_oph_bw}, \ref{fig:xx_vir_bw} and \ref{fig:uy_eri_bw} show the relation between the angular size of the star in phases corresponding to the $K_s$ band measurements, and the integrated radial velocity in corresponding phases, multiplied by $2 \omega /1AU$ to have identical units (mas) for both axes. The slope of this relation is simply the projection factor and the zero$-$point is the mean angular diameter of the star which is used to calculate the mean radius. We fit the line using the least squares method. Our results for V716~Oph, UY~Eri and XX~Vir for different SBCRs are presented in Table \ref{tab:pfactor_results}. Statistical uncertainties are evaluated using the Monte Carlo simulations. In each simulation we generate artificial light and radial velocity curves from the normal distributions defined by original measurements and corresponding uncertainties and perform the BW analysis. We fit the Gaussian profiles to the resulting histograms of p$-$factor and radius to obtain the final value of the given parameter and its uncertainty. 

The obtained values of projection factors for different SBCRs for each star agree within 1$\sigma$ statistical uncertainties. In the case of radii we observe a systematic shift of radii depending on the SBCR used to estimate the angular diameter. Similarly as in \citet{2024AA...689A.241W}, the relation from \citet{2004AA...426..297K} is considered as the reference. The uncertainty of p$-$factor and radius of a given star, resulting from the uncertainty of the surface brightness is calculated as the standard deviation from the mean value of a given parameter and results are presented in Columns 3 and 10 of Table \ref{tab:errors}. In the same Table we present the systematic errors on p$-$factors and radii related to the systematic uncertainty of $V$ and $K_{\mathrm{S}}$ light curves, reddening and parallax. We estimated these errors by shifting the given curve or parameter by the corresponding systematic error value and repeating the BW analysis. The difference between the new p$-$factor or mean radius and the original value is considered as the systematic uncertainty. In columns 8 and 15 we give the total errors on p$-$factors and radii, respectively, calculated as the quadratic sum of errors.

In the case of V716~Oph and UY~Eri, the obtained projection factors are between the mean values obtained in \citet{2024AA...689A.241W} and \citet{2024AA...690A.295Z} using an identical technique for Type II Cepheids and RR Lyrae, which (for the relation $K04b$) amount to 1.330$\pm$0.058 and 1.398$\pm$0.078, respectively. 

The calculated projection factor for XX~Vir is very high. We suspect that this is the result of the underestimated parallax in the $Gaia$ catalogue as its absolute magnitude in Figure \ref{fig:mw_acep_plr}, and radius as is shown below, are also quite significantly higher than expected for ACeps of similar period. If we assume the value of the projection factor for this star to be 1.35, the BW method described above gives the the parallax value of 0.21\,mas. When we adopt this value to calculate the distance modulus of XX~Vir, its absolute magnitude and mean radius are much closer to the corresponding relations. 


To our knowledge there are no projection factor measurements for stars clearly classified as ACeps in the literature. However \citet{2017ApJ...842..110P} analyzed the LMC Type II Cepheid OGLE$-$LMC$-$T2CEP$-$098 in a binary system and it turned out that the pulsating component has a mass of $\sim$1.5\,$M_{\odot}$, which is expected for ACeps. In the PL diagram it lies on the extension of the relation for ACeps towards longer periods (period of the pulsating component is 4.974\,d). The projection factor was estimated by the authors to be 1.30$\pm$0.03, which is lower but in 1$\sigma$ agreement with the values measured in this work for V716~Oph and UY~Eri. It is however important to note, that the methodology used in \citet{2017ApJ...842..110P} is different from the present approach and uses pulsation theory. It is known that the projection factor is method dependent \citep[see e.g.][and references therein]{2008AA...489.1255N,2019AA...631A..37B,2024AA...689A.241W}. The values of p$-$factors obtained for V716~Oph and UY~Eri are also consistent with a theoretical prediction by \citet{2012A&A...541A.134N} based on a simple plane$-$parallel and spherically$-$symmetric model of stellar atmospheres of short period pulsating stars. This suggest that velocity gradients in the stellar atmosphere have low impact on the p$-$factor in short period regime.

\begin{figure}[]
            \centering
            \includegraphics[width=0.5\textwidth]{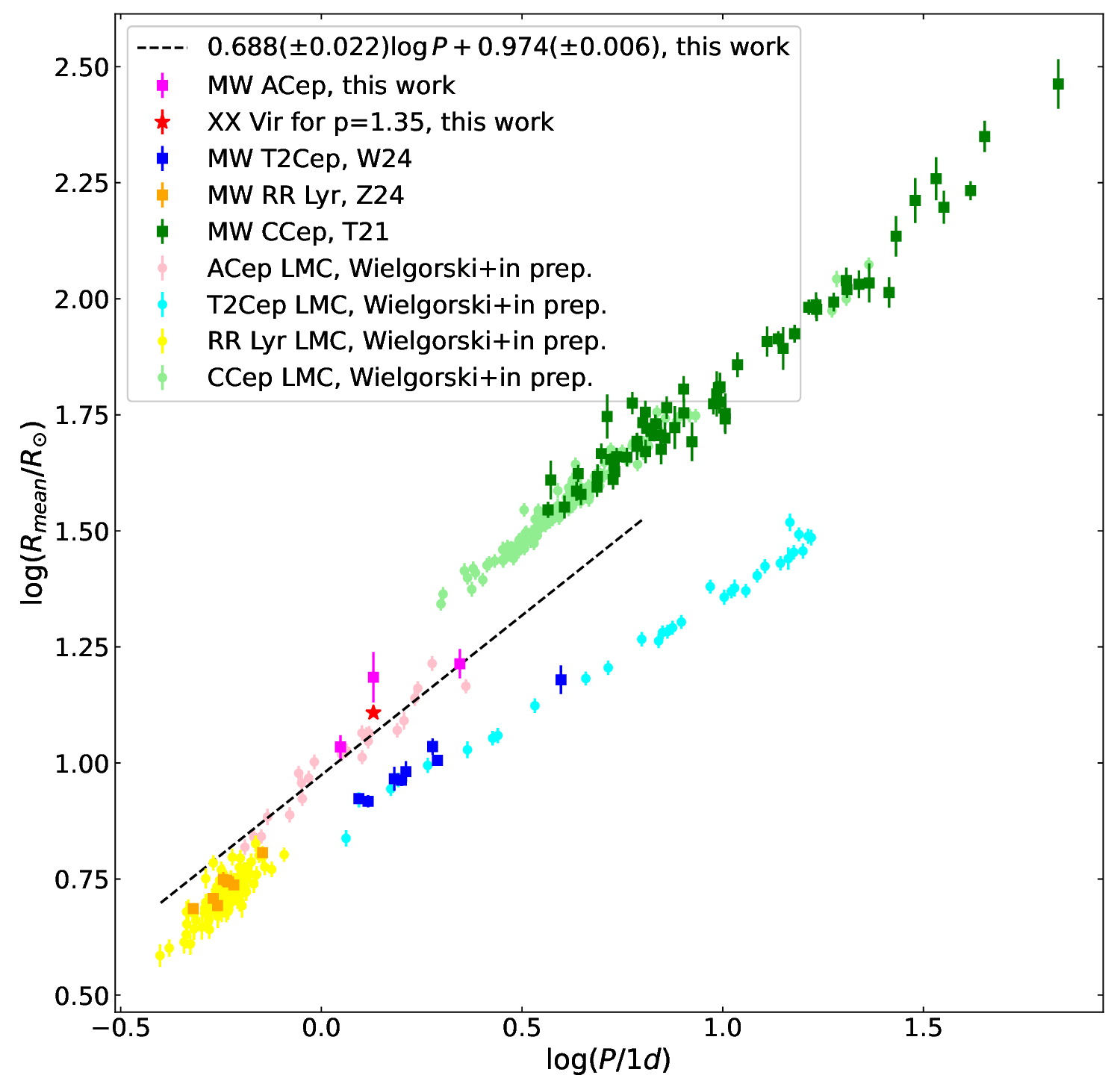}
            \caption{Period$-$radius diagram for three Anomalous Cepheids analysed in this work, Milky Way Type II Cepheids from \citet[][W24]{2024AA...689A.241W}, RR Lyrae from \citet[][Z24]{2024AA...690A.295Z} and Classical Cepheids from \citet[][T21]{2021AA...656A.102T}, and LMC Classical, Anomalous and Type II Cepheids and RR Lyrae stars from Wielg\'orski et al. (in prep.). Black dashed line is the linear fit for the LMC ACeps.\label{fig:prr}}
            \end{figure}

In Figure \ref{fig:prr} the measured radii of the three analysed ACeps are plotted against their pulsational periods. Period$-$radius relations of MW Type II Cepheids, RR Lyrae and Classical Cepheids from \citet{2024AA...689A.241W}, \citet{2024AA...690A.295Z} and \citet{2021AA...656A.102T} are also plotted for comparison. Additionally, we plot the mean radii of LMC Anomalous, Classical and Type II Cepheids and RR Lyrae type stars from Wielg\'orski et al. (in prep.) estimated from OGLE $V$ passband and VMC $K_\mathrm{S}$ passband photometry, K04b SBCR and the distance from eclipsing binaries. The measured radii of V716~Oph and UY~Eri are located well on the period$-$radius relation formed by the LMC ACeps. The radius of XX~Vir is significantly bigger than radii of the LMC ACeps with similar periods. As discussed above, we suspect that the parallax of this star is underestimated in $Gaia$ DR3 and assuming the projection factor value of 1.35 from BW analysis we obtained the radius of 12.96$\pm$0.04(stat)\,$R_{\odot}$, which is much closer to the period$-$radius relation (red star in Figure \ref{fig:prr}). The radius of this star was estimated in the BW analysis based on solely optical photometry by \citet{1997AJ....113.1833B}. The authors assumed the value of the p$-$factor to be 1.38 from \citet{1989ApJ...342..467G}, and they obtained a radius of 13$R_{\odot}$ but the uncertainty of this measurement was 7\,$R_{\odot}$.

 The black dashed line in Figure \ref{fig:prr} is the linear fit to the LMC ACeps data:
 \begin{equation}
 \log \langle R \rangle=0.688 \pm 0.022 \log P + 0.974 \pm 0.006.
 \end{equation}
 The radii of V716~Oph, UY~Eri, and XX~Vir, assuming the p$-$factor value of 1.35, are in a good agreement with this relation. It is worth to note that the agreement of the mean radius obtained in the BW analysis of the pulsating star with the respective period-radius relation is an important additional information which validates the distances measured with this technique. The slope and zero$-$point of the obtained period-radius relation are in perfect agreement with the one found by \citet{2017A&A...604A..29G} based on Spectral Energy Distribution (SED) fitting of ACeps in LMC:
 \begin{equation}
 \log \langle R \rangle=0.692 \pm 0.034 \log P + 0.972 \pm 0.005.
 \end{equation}
  
\begin{table}[h]
    \caption{Projection factors and mean radii obtained for the analysed Anomalous Cepheids for different Surface Brightness$-$Color Relations.}
    \label{tab:pfactor_results}
    \centering
    \begin{tabular}{cccc}
    \hline\hline
   
    $Name$ & $SBCR$ & $p$ & $\langle R \rangle$\\
     &  & &  ($R_\odot$)  \\
    \hline
    \hline
    V716 Oph & K04a & 1.36$\pm$0.09 & 10.88$\pm$0.02\\
    & \textbf{K04b} & \textbf{1.38$\pm$0.10} & \textbf{10.67$\pm$0.03}\\
    & S21 & 1.42$\pm$0.14 & 10.60$\pm$0.03\\
    & G21 & 1.36$\pm$0.12 & 10.65$\pm$0.02\\
    & B25 & 1.35$\pm$0.10 & 10.74$\pm$0.03\\
    \hline
    XX Vir & K04a & 1.53$\pm$0.07 & 15.50$\pm$0.03\\
    & \textbf{K04b} & \textbf{1.59$\pm$0.06} & \textbf{15.20$\pm$0.03}\\
    & S21 & 1.63$\pm$0.09 & 15.10$\pm$0.04\\
    & G21 & 1.56$\pm$0.07 & 15.18$\pm$0.03\\
    & B25 & 1.52$\pm$0.06 & 15.30$\pm$0.03\\
    \hline
    UY Eri & K04a & 1.35$\pm$0.11 & 16.60$\pm$0.04\\
    & \textbf{K04b} & \textbf{1.35$\pm$0.10} & \textbf{16.37$\pm$0.04}\\
    & S21 & 1.35$\pm$0.12 & 16.42$\pm$0.05\\
    & G21 & 1.36$\pm$0.11 & 16.42$\pm$0.04\\
    & B25 & 1.34$\pm$0.09 & 16.39$\pm$0.03\\
    \hline
   
    \hline
    \end{tabular}
    \tablefoot{The presented uncertainties are the statistical errors of the BW fit calculated in the Monte$-$Carlo simulations. Total errors are given in Table \ref{tab:errors}}
    \end{table}

    \begin{table*}
        \caption{Summary of the errors obtained for p$-$factors and radii for each Anomalous Cepheid.}
        \label{tab:errors}
        \centering
        \begin{tabular}{c|ccccccc|ccccccc}
        \hline\hline
       
        $Name$ & $\sigma_{stat}$ & $\sigma_{SBCR}$ &  $\sigma_{V}$ & $\sigma_{K_{\mathrm{S}}}$ & $\sigma_{E(B-V)}$ & $\sigma_{\omega}$  & \textbf{$\sigma_{total}$} & $\sigma_{stat}$ & $\sigma_{SBCR}$ & $\sigma_{V}$ & $\sigma_{K_{\mathrm{S}}}$ & $\sigma_{E(B-V)}$ & $\sigma_{\omega}$  & \textbf{$\sigma_{total}$}\\
        \hline
        & \multicolumn{7}{c}{$p$-factors} & \multicolumn{7}{c}{$\langle R \rangle$ ($R_{\odot}$)}\\
        \hline
        V716 Oph & 0.10 & 0.02	& 0.01	& 0.02	& 0.01	& 0.08 & \textbf{0.13} & 0.03 & 0.10	& 0.03 & 0.14 & 0.06 & 0.60 & \textbf{0.63}\\
        XX Vir  & 0.06 & 0.04	& 0.01	& 0.02	& 0.01	& 0.20 & \textbf{0.21} & 0.03 & 0.14	& 0.03 & 0.19 & 0.10 & 1.90 & \textbf{1.92}\\
        UY Eri & 0.10 & 0.01	& 0.01	& 0.02	& 0.01	& 0.10 & \textbf{0.14} & 0.04 & 0.08	& 0.06 & 0.20 & 0.10 & 1.16 & \textbf{1.19}\\
        
        \hline
        \end{tabular}
        \tablefoot{ The total errors $\sigma_{total}$ are the quadratic sum of the statistical errors of the BW fit $\sigma_{stat}$, errors related to the uncertainty of the surface brightness $\sigma_{SBCR}$, errors related to the systematic shift of the $V$ passband light curve $\sigma_{V}$, errors related to the systematic shift of the $K_{\mathrm{S}}$ band light curve $\sigma_{K_{\mathrm{S}}}$, errors related to the $E(B-V)$ uncertainty $\sigma_{E(B-V)}$, and errors related to the uncertainty of the parallax $\sigma_{\omega}$.}
        \end{table*}

\section{Summary}\label{sec:summary}

Using new photometry for a sample of Anomalous Cepheids from the solar neighbourhood in the Johnson $B$, $V$, 2MASS $J$, $H$, $K_{\mathrm{S}}$ and Pan$-$STARRS $g$, $r$, $i$ passbands, and parallaxes from the $Gaia$ DR3 we determine zero$-$points of the Period$-$Luminosity and Period$-$Wesenheit Relations, assuming the literature slopes measured for LMC ACeps. The difference between the zero$-$points of PLRs in the MW and in the LMC is used to estimate the true distance modulus of the LMC. For $W_{JK}$ Wesenheit index we obtained the distance modulus of 18.454$\pm$0.045\,mag, in a very good agreement with the  distance modulus from eclipsing binaries \citep{2019Natur.567..200P}. The systematic uncertainty of our zero$-$points is estimated to be about 0.1mag. 

We used our photometry in $V$ and $K_{\mathrm{S}}$ and new spectroscopic data to perform the Baade$-$Wesselink analysis of three nearby Anomalous Cepheids with parallaxes measured by $Gaia$, aiming in determining projection factors of these stars. For V716~Oph and UY~Eri we obtained projection factors of 1.38$\pm$0.13 and 1.34$\pm$0.14, respectively. For XX~Vir the obtained projection factor is very high (1.59$\pm$0.21). We suspect that the parallax of this star is underestimated in the $Gaia$ DR3 as the absolute magnitude of the star and its radius obtained in this study are higher than expected for Anomalous Cepheids of similar periods. Assuming the projection factor of this star to be 1.35, we measured the distance of this star from the Baade$-$Wesselink method; the new absolute brightness and mean radius agree with the PLR and period$-$radius relation of LMC ACeps. 

We expect that $Gaia$ parallaxes will be more accurate in the next Data Release, which will allow to increase the precision of PLRs, p$-$factors and radii. We moreover plan to expand the sample of Anomalous Cepheids to be observed from the OCM and to recalibrate both techniques in the future.

\begin{center}
ACKNOWLEDGMENTS
\end{center}

We thank the anonymous referee for very valuable comments and suggestions, which helped us to improve the manuscript.
The research leading to these results has received funding from the European Research Council (ERC) under the European Union’s Horizon 2020 research and innovation programme (grant agreements No 695099 and No 951549). Support from DIR/2024/WK/02 grant of the Polish Ministry of Science and Higher Education and the Polish National Science Center grants MAESTRO 2017/26/A/ST9/00446 and BEETHOVEN 2018/31/G/ST9/03050 and Polish$-$French Marie Skłodowska$-$Curie and Pierre Curie Science Prize awarded by the Foundation for Polish Science is also acknowledged. W.G. and G.P. gratefully acknowledge financial support for this work from the BASAL Centro de Astrofisica y Tecnologias Afines (CATA) AFB$-$170002. N.N. acknowledges the support of the French Agence Nationale de la Recherche (ANR) under grant ANR$-$23$-$CE31$-$0009$-$01 (Unlock$-$pfactor). B.P. acknowledges support from the Polish National Science Center grant SONATA BIS 2020/38/E/ST9/00486. R.Sm. acknowledges support from  the National Science Center, Poland, Sonata BIS project 2018/30/E/ST9/00598. F.P. gratefully acknowledges the generous and invaluable support of the Klaus Tschira Foundation. Development of the IRIS camera was supported by the National Science Foundation under grant AST 0704954.
    
Based on data collected under the ESO/CAMK PAN – USB agreement at the ESO Paranal Observatory. We thank our colleagues involved in the process of gathering data at Rolf Chini Cerro Murphy Observatory. 
    
Based on observations collected at the European Organisation for Astronomical Research in the Southern Hemisphere under ESO programmes 099.D$-$0380(A), 0100.D$-$0339(B),0100.D$-$0273(A),0102.D$-$0281(A), 105.20L8.002, 106.20Z1.001, 106.20Z1.002, 106.21T1.001, 108.22JX.001, 111.24YL.001 and CNTAC programmes CN2016B$-$150, CN2017A$-$121, CN2017B$-$43, CN2018A$-$40, CN2019B$-$64, CN2020B$-$42, CN2020B$-$69. We are greatly indebted to the staff at the ESO La Silla and Paranal Observatories for excellent support during the many visitor mode runs and for performing the observations in the service mode.
    
This work has made use of data from the European Space Agency (ESA) mission
    {\it Gaia} (\url{https://www.cosmos.esa.int/gaia}), processed by the {\it Gaia}
    Data Processing and Analysis Consortium (DPAC,
    \url{https://www.cosmos.esa.int/web/gaia/dpac/consortium}). Funding for the DPAC
    has been provided by national institutions, in particular the institutions
    participating in the {\it Gaia} Multilateral Agreement.
    
This publication makes use of data products from the Two Micron All Sky Survey \citep{2006AJ....131.1163S}, which is a joint project of the University of Massachusetts and the Infrared Processing and Analysis Center/California Institute of Technology, funded by the National Aeronautics and Space Administration and the National Science Foundation. This research has made use of the SIMBAD database, operated at CDS, Strasbourg, France \citep{2000A&AS..143....9W}. We acknowledge with thanks the variable star observations from the AAVSO International Database contributed by observers worldwide and used in this research.

Software used in this work: \texttt{gaiadr3$\_$zero$-$point} \citep{2021A&A...649A...4L}, \texttt{Astropy7} \citep{2013A&A...558A..33A,2018AJ....156..123A}, \texttt{IRAF} \citep{1986SPIE..627..733T,1993ASPC...52..173T}, \texttt{Sextractor} \citep{1996A&AS..117..393B}, \texttt{SCAMP} \citep{2006ASPC..351..112B}, \texttt{SWARP} \citep{2010ascl.soft10068B}, \texttt{DAOPHOT} \citep{1987PASP...99..191S}, \texttt{NumPy} \citep{2011CSE....13b..22V}, \texttt{SciPy} \citep{2020SciPy-NMeth}, \texttt{Matplotlib} \citep{2007CSE.....9...90H}, \texttt{Ravespan} \citep{2017ApJ...842..110P}. The custom software with GUI used for the Baade$-$Wesselink analysis is available on Github: \url{https://github.com/araucaria-project/balwan.git}.

\bibliography{acep}{}
\bibliographystyle{aa}

\begin{appendix}
    \section{Figures with light curves and the BW analysis}\label{ap:fig}

\begin{figure*}[]
    \centering
    \includegraphics[width=0.8\textwidth]{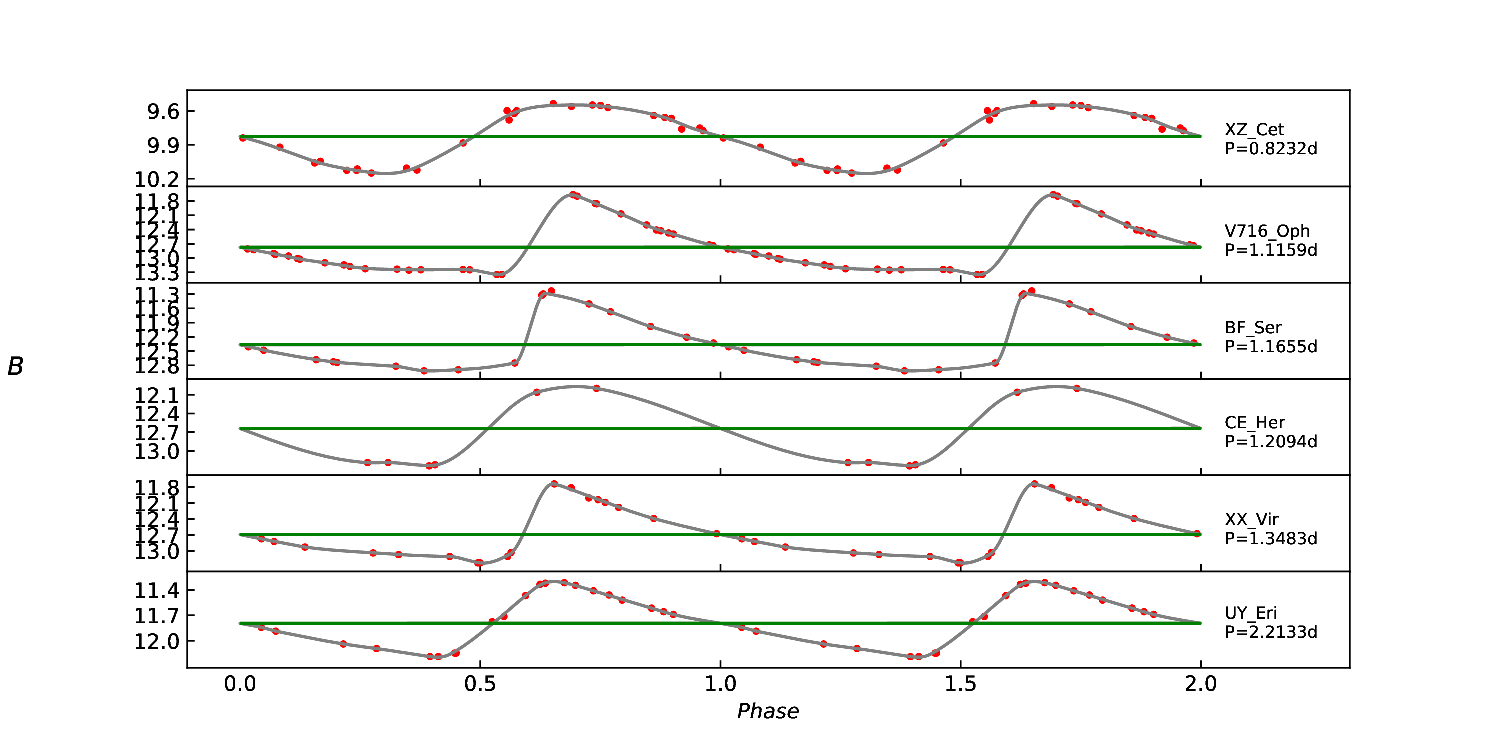}
    \caption{$B$ light curves. Green line is the measured mean magnitude.\label{fig:v16_b}}
    \end{figure*}

        \begin{figure*}[]
        \centering
        \includegraphics[width=0.8\textwidth]{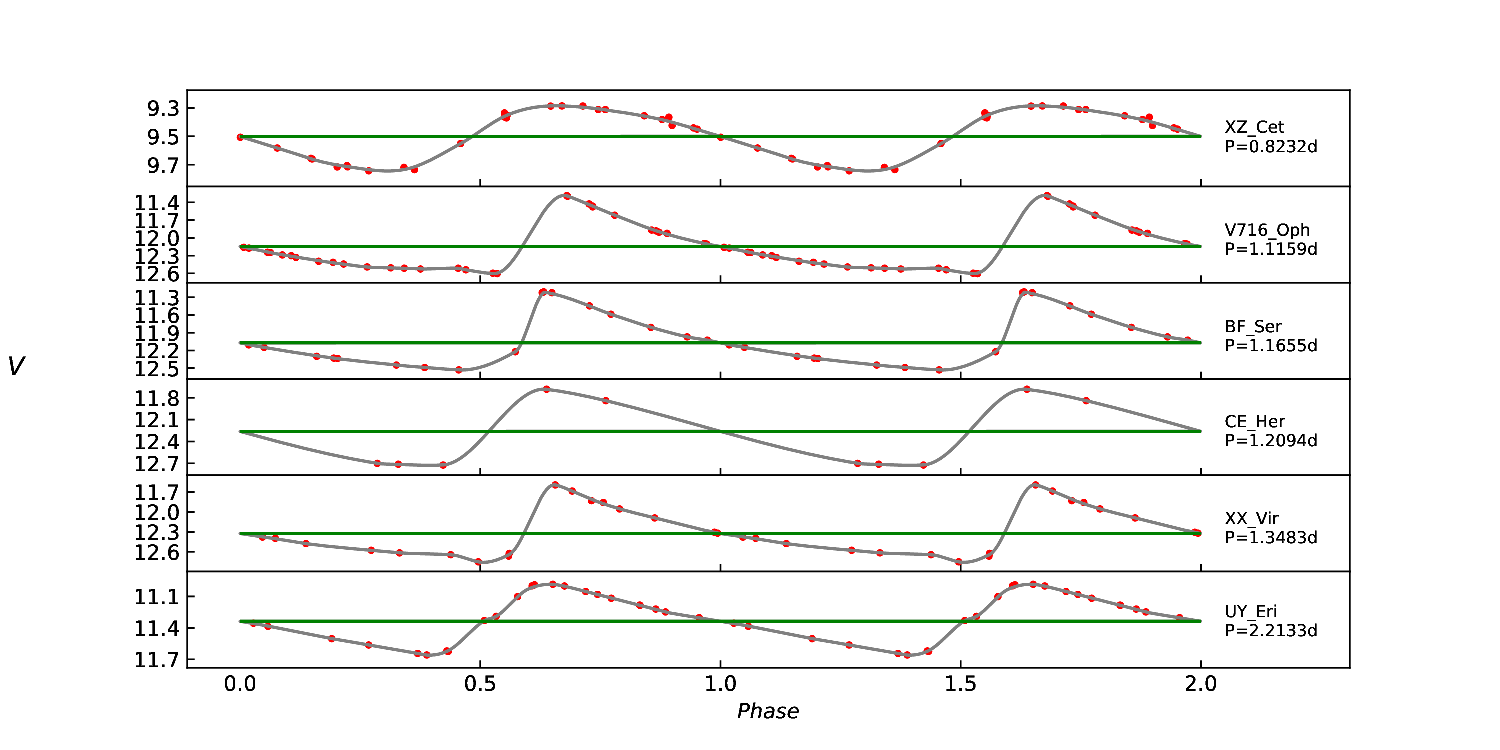}
        \caption{$V$ light curves. Green line is the measured mean magnitude.\label{fig:v16_v}}
        \end{figure*}

\begin{figure*}[]
\centering
\includegraphics[width=0.8\textwidth]{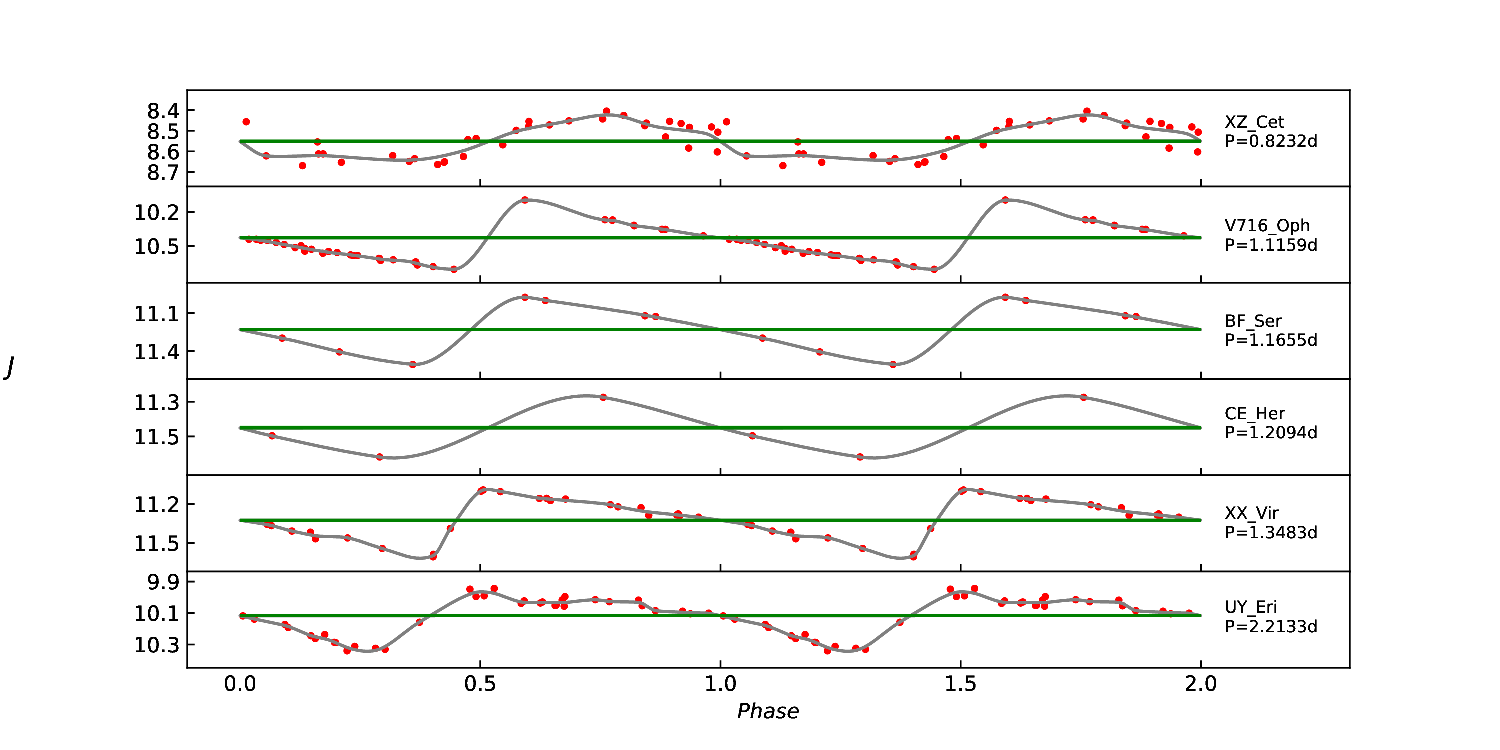}
\caption{$J$ light curves. Green line is the measured mean magnitude.\label{fig:j}}
\end{figure*}

\begin{figure*}[]
\centering
\includegraphics[width=0.8\textwidth]{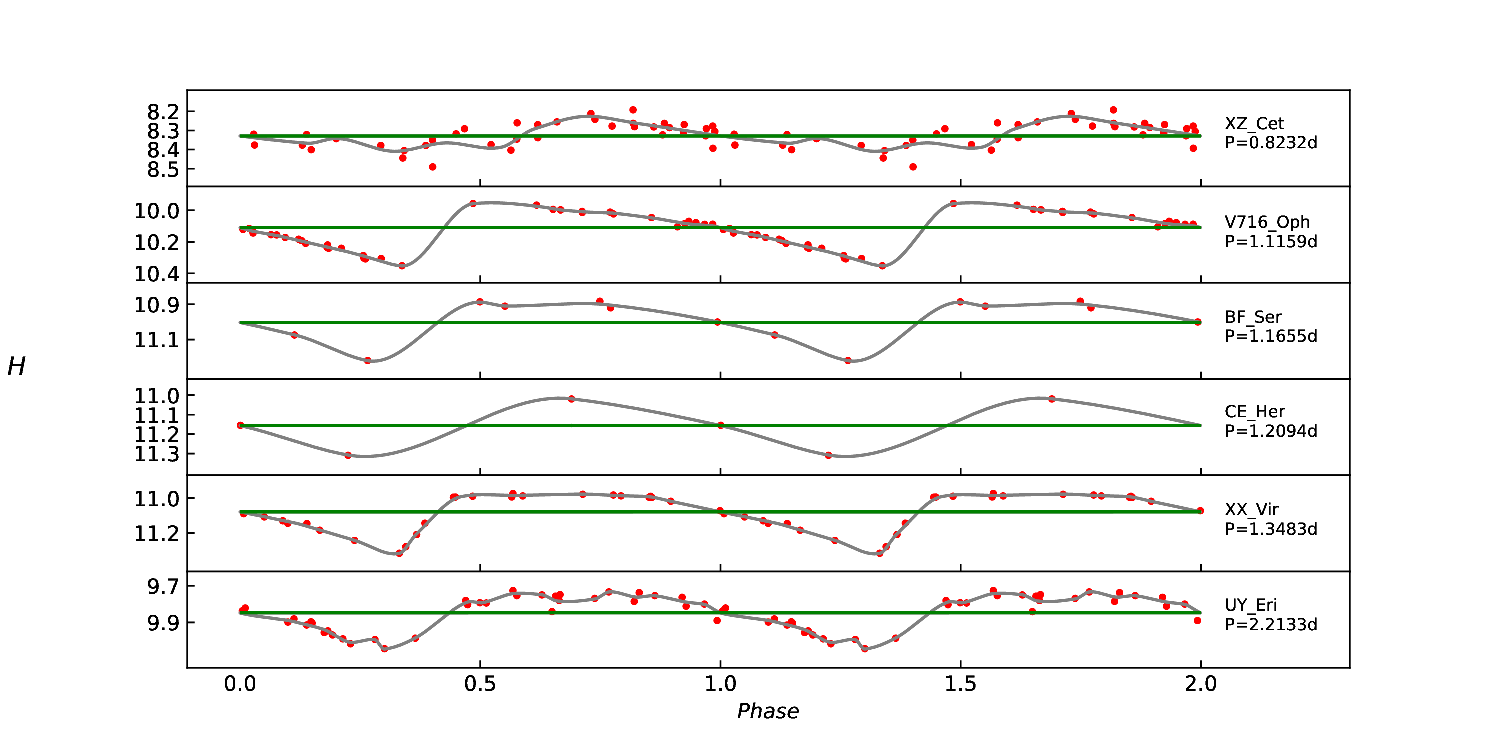}
\caption{$H$ light curves. Green line is the measured mean magnitude.\label{fig:h}}
\end{figure*}

\begin{figure*}[]
\centering
\includegraphics[width=0.8\textwidth]{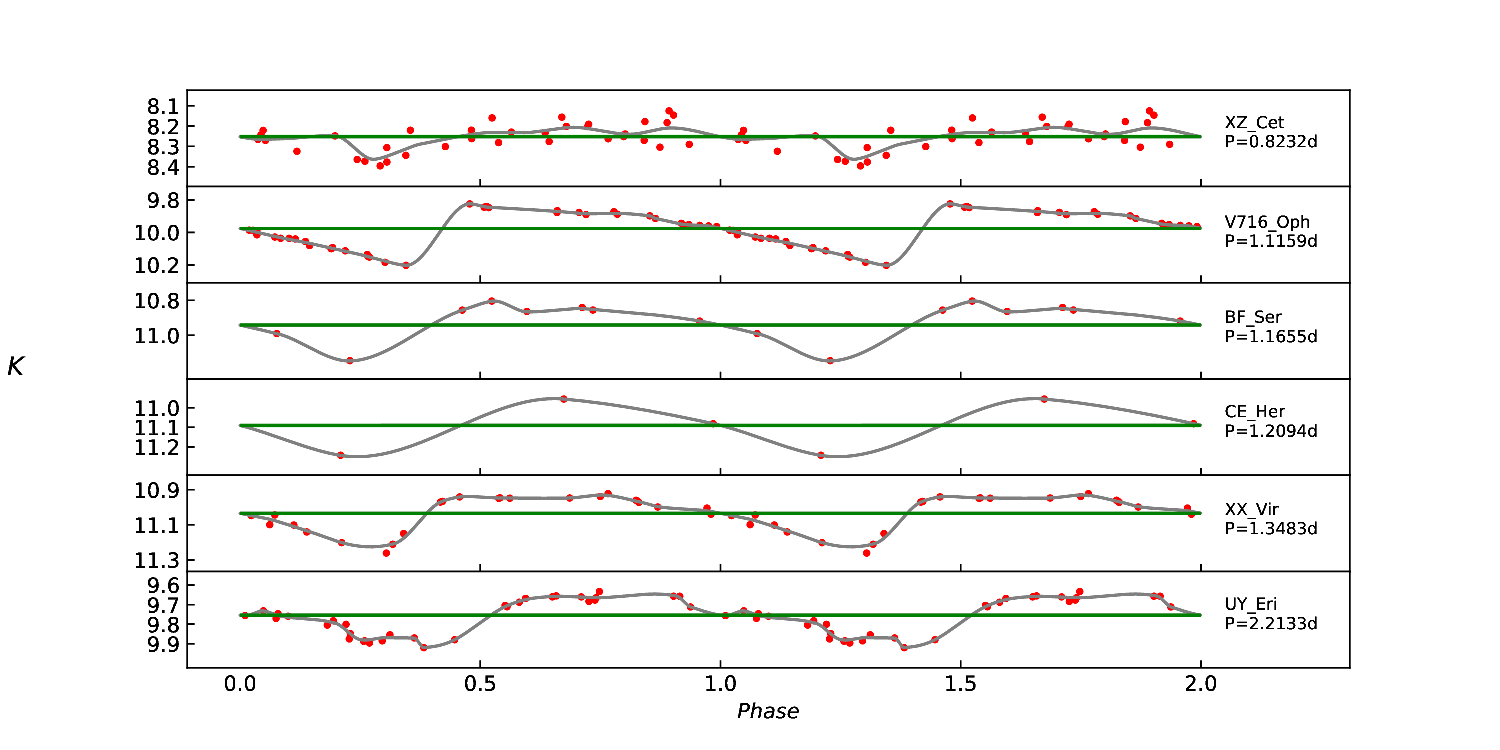}
\caption{$K_{\mathrm{S}}$ light curves. Green line is the measured mean magnitude.\label{fig:k}}
\end{figure*}

\begin{figure*}[]
    \centering
    \includegraphics[width=0.8\textwidth]{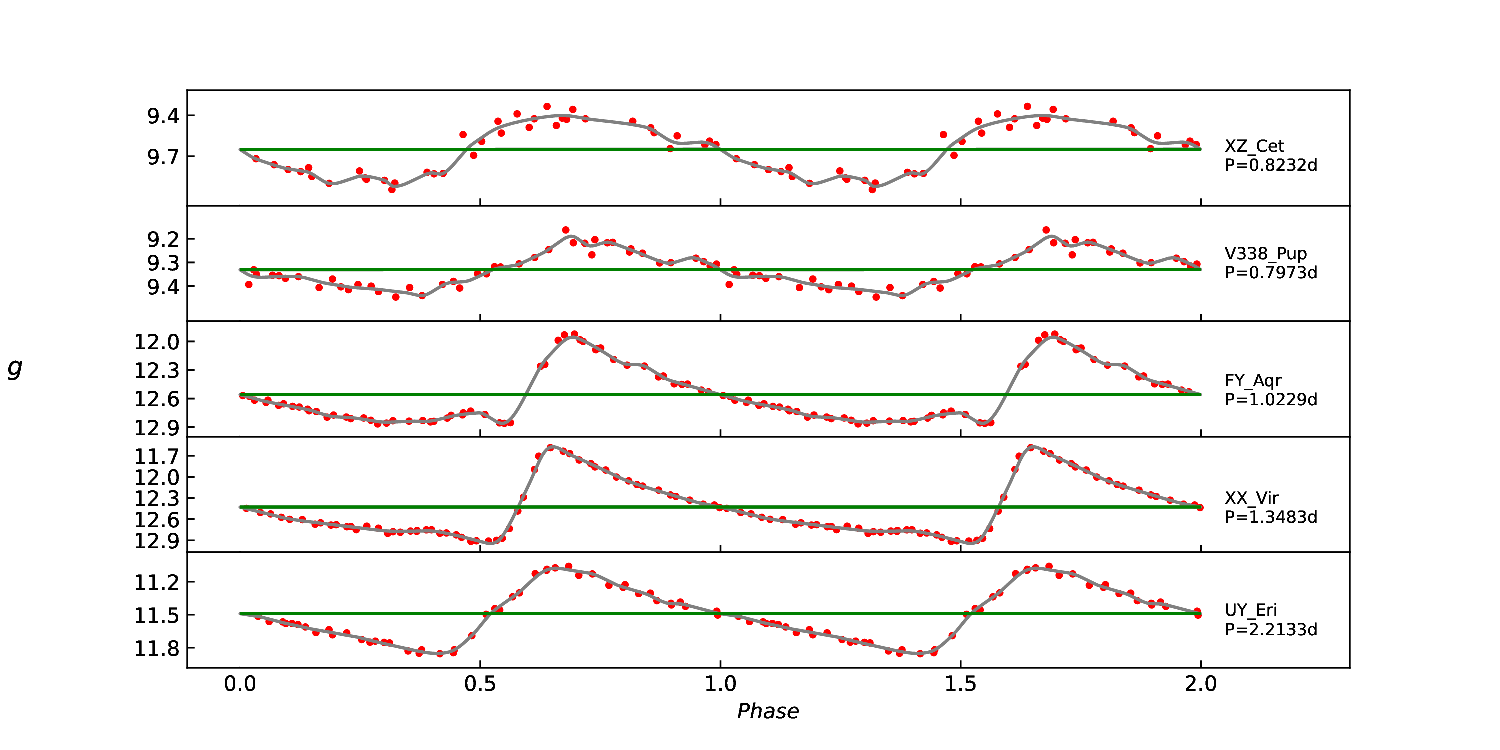}
    \caption{Pan$-$STARRS $g$ light curves. Green line is the measured mean magnitude.\label{fig:g}}
    \end{figure*}

    \begin{figure*}[]
        \centering
        \includegraphics[width=0.8\textwidth]{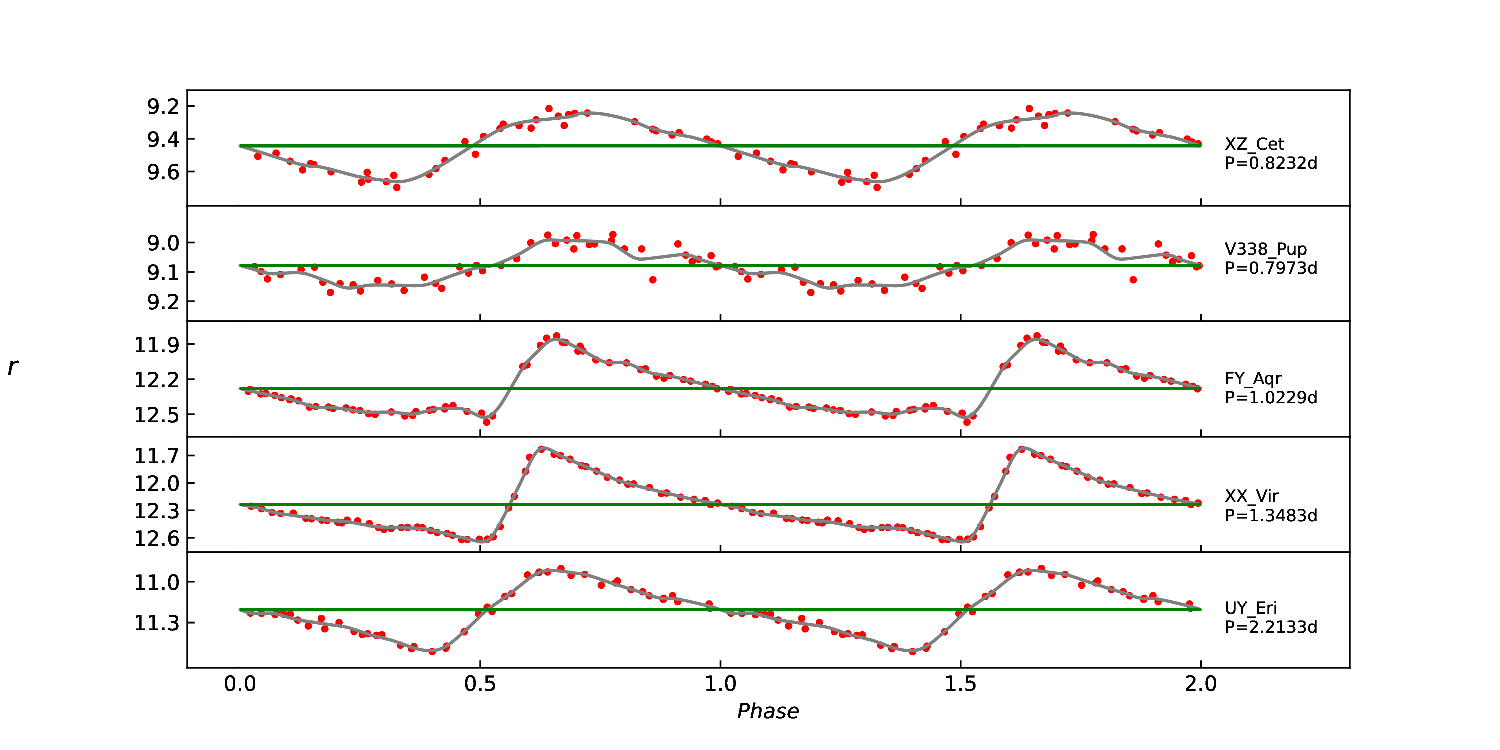}
        \caption{Pan$-$STARRS $r$ light curves. Green line is the measured mean magnitude.\label{fig:r}}
        \end{figure*}

        \begin{figure*}[]
            \centering
            \includegraphics[width=0.8\textwidth]{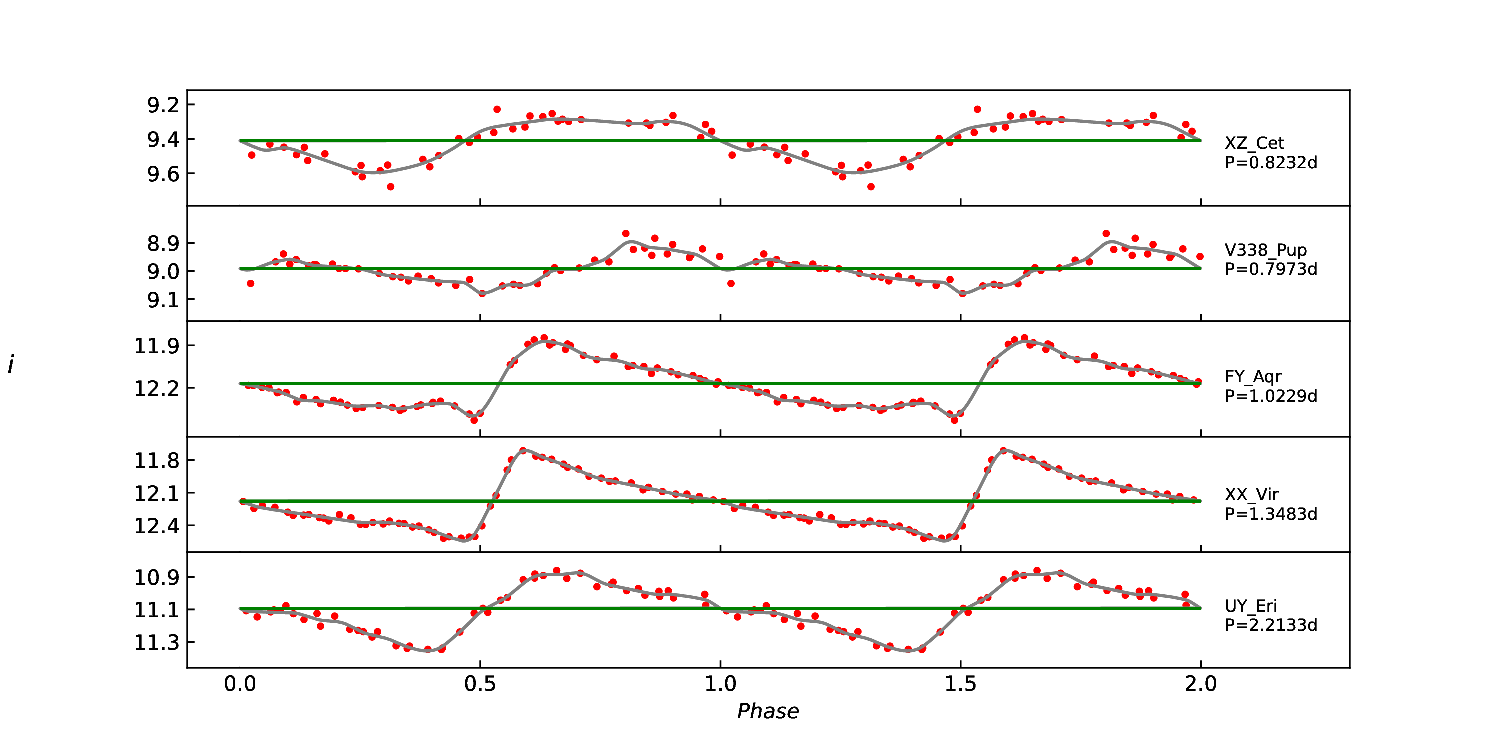}
            \caption{Pan$-$STARRS $i$ light curves. Green line is the measured mean magnitude.\label{fig:i}}
            \end{figure*}

\begin{figure*}[]
\centering
\includegraphics[width=0.8\textwidth]{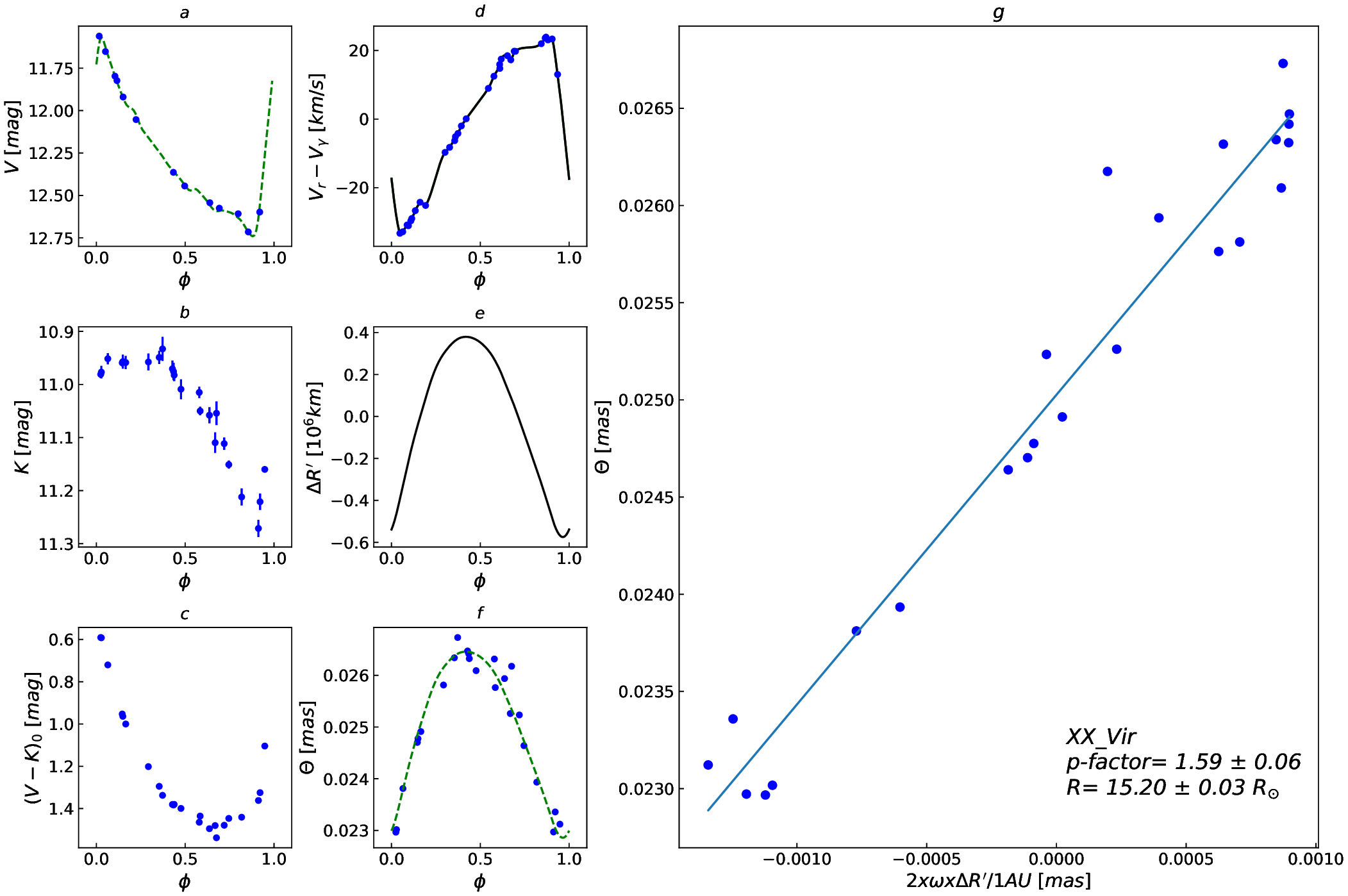}
\caption{The Baade$-$Wesselink analysis of XX Vir. For a description of the panels, see Figure \ref{fig:v716_oph_bw}.\label{fig:xx_vir_bw}}
\end{figure*}

\begin{figure*}[]
\centering
\includegraphics[width=0.8\textwidth]{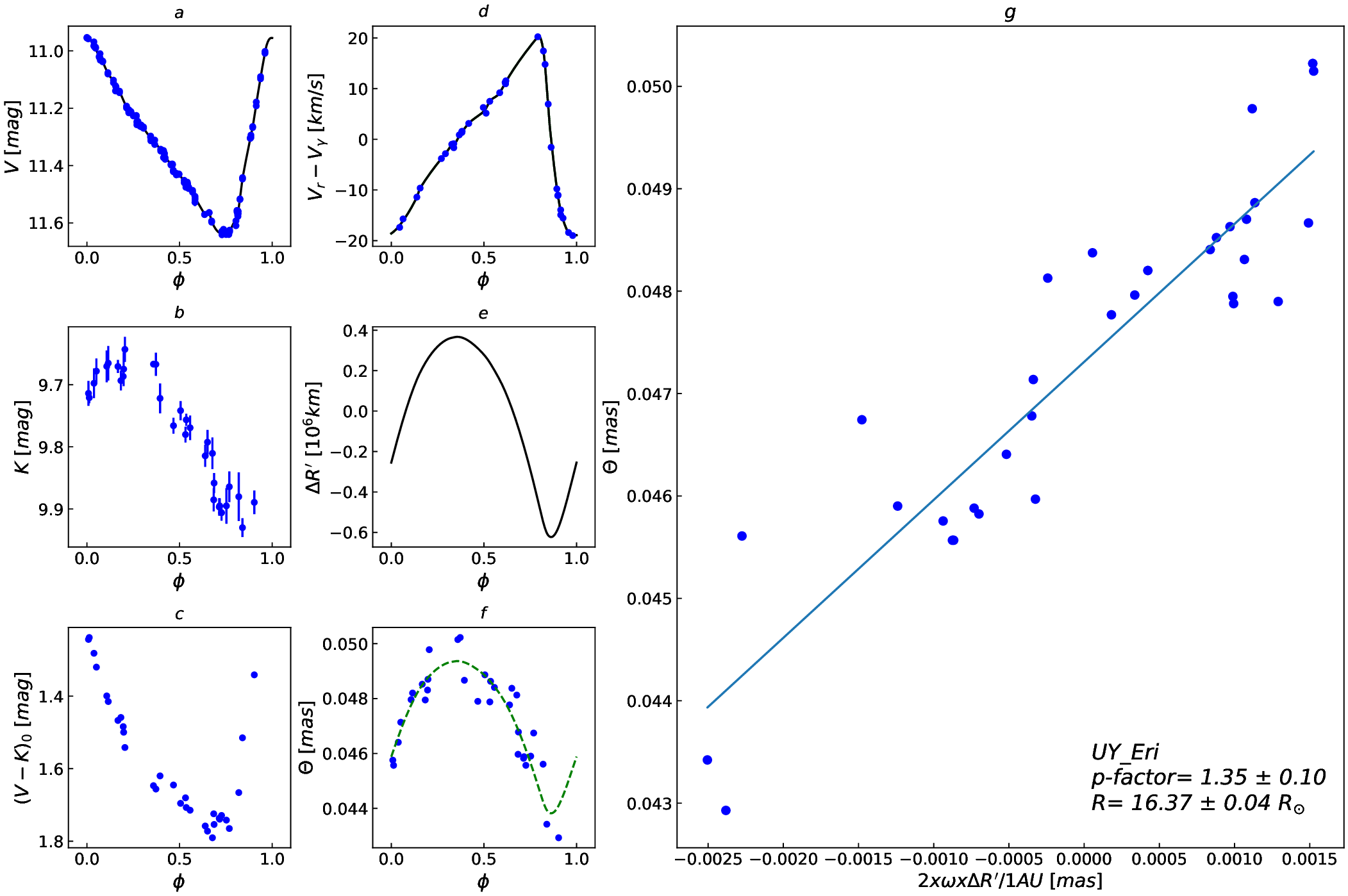}
\caption{The Baade$-$Wesselink analysis of UY Eri. For a description of the panels, see Figure \ref{fig:v716_oph_bw}. \label{fig:uy_eri_bw}}
\end{figure*}

\clearpage


\end{appendix}

\end{document}